\documentclass[
  reprint,
  nofootinbib,
  amsmath,amssymb,
  aps,
  prd,
  floatfix,
]{revtex4-1}

\usepackage[USenglish]{babel}

\usepackage{amsfonts}
\usepackage{amsbsy}
\usepackage{bm}
\usepackage{mathrsfs}
\usepackage{slashed}
\usepackage{extarrows}

\usepackage{graphicx}
\usepackage{epstopdf}
\usepackage{dcolumn}

\usepackage[dvipsnames]{xcolor}

\usepackage{array}
\usepackage{multirow}
\usepackage{enumitem}

\usepackage[caption=false]{subfig}

\usepackage{url}
\usepackage{varioref}
\usepackage{float}
\usepackage{verbatim}
\usepackage{latexsym}
\usepackage{orcidlink}
\usepackage{setspace}
\usepackage[normalem]{ulem}

\PassOptionsToPackage{
    colorlinks,
    citecolor=blue,
    urlcolor=magenta,
    linkcolor=blue
}{hyperref}
\usepackage{hyperref}

\newcommand{\GSSI}{Gran Sasso Science Institute (GSSI), I-67100 L’Aquila, Italy}
\newcommand{\GranSasso}{INFN, Laboratori Nazionali del Gran Sasso, I-67100 Assergi, Italy}

\newcommand{\IUCAA}{Inter-University Centre for Astronomy and Astrophysics\\ PostBag 4, Ganeshkhind, Pune-411007, Maharashtra, India}

\begin{document}

\title{Geometric properties of slowly rotating black holes embedded in matter environments}


\author{Sayak Datta\orcidlink{0000-0002-4774-0298}}
\email{sayak.datta@gssi.it}
\address{\GSSI}
\address{\GranSasso}

\author{Chiranjeeb Singha \orcidlink{0000-0003-0441-318X}}
\email{chiranjeeb.singha@iucaa.in}
\address{\IUCAA}


\begin{abstract} 
Extreme mass-ratio inspirals (EMRIs) provide a uniquely precise probe of strong-field gravity, where small deviations from vacuum Kerr geometry can accumulate over many orbital cycles. In realistic astrophysical settings, black holes are embedded in surrounding dark and baryonic matter whose presence and motion can perturb the spacetime.
A systematic semi-analytic framework incorporating the rotation of the environment itself into a black-hole geometry, and propagating its effects consistently into conserved quantities and epicyclic observables, has not been explored exhaustively, limiting consistent assessments of environmental effects on precision observables.
In this work, we construct a slowly rotating black hole spacetime embedded in an anisotropic matter distribution and explicitly include the angular velocity of the surrounding medium within a controlled slow-rotation expansion. We demonstrate that the environment’s velocity field induces corrections to the metric coefficients that propagate into modifications of conserved quantities governing geodesics. We also derive semi-analytic shifts in the innermost stable circular orbit, light-ring location, and radial and vertical epicyclic frequencies, showing that environmental rotation produces systematic and in certain regimes qualitatively distinct behavior relative to static configurations. Consequently, we explicitly show that the environment's nature and motion shift the positions of the epicyclic resonances.
The formalism applies to generic anisotropic matter profiles and is not restricted to the specific halo model adopted for numerical illustration.
These results establish a direct and quantitatively controlled link between environmental rotation and strong-field orbital observables, enabling consistent incorporation of rotating matter environments into precision EMRI modeling.
\end{abstract}

\maketitle

\section{Introduction}

Extreme mass-ratio inspirals (EMRIs) and other precision strong-field probes are expected to measure black hole spacetimes with unprecedented accuracy, where even small deviations from the Kerr geometry can accumulate into observable phase shifts over many orbital cycles. The Schwarzschild and Kerr metrics, in particular, hence play a central role in EMRI waveform modeling.
In realistic astrophysical settings, however, black holes are not truly isolated. Instead, they are generically embedded in complex, matter-rich environments, including accretion disks, stellar clusters, gas clouds, and extended dark matter (DM) halos that dominate the mass distribution on galactic and subgalactic scales \cite{Barausse:2014tra,BinneyTremaine,Yunes:2011ws,Cardoso:2019rou,Cardoso:2020iji,Derdzinski:2020wlw,Zwick:2022dih,Cardoso:2021wlq}. The presence of such surrounding media can significantly distort the local spacetime geometry, modify geodesic motion, shift characteristic orbital radii, and imprint potentially observable signatures on electromagnetic and gravitational-wave signals. In particular, environmental effects may encode valuable information about the microscopic properties and macroscopic distribution of the ambient matter. A consistent modeling of such environmental effects is therefore essential for reliable interpretation of high-precision gravitational-wave observations.

These considerations are particularly relevant for EMRIs, in which a stellar-mass compact object spirals into a supermassive black hole. Future space-based detectors such as \emph{LISA} \cite{2017arXiv170200786A} or \emph{TianQin} \cite{TianQin:2015yph} will measure the near-horizon geometry with high precision, so even small deviations from vacuum Kerr geometry may accumulate into observable phase shifts. Accurately modeling environmental effects is therefore essential for robust parameter estimation. 
A systematic investigation of such effects requires solving the Einstein field equations in the presence of a nontrivial stress–energy tensor that faithfully captures the physical characteristics of the surrounding matter distribution, both baryonic and dark type. Among the various possible models to describe environment, the example case of DM halos, as the \emph{Einstein cluster} \cite{7bb06a79-8225-31c6-88c3-0c4f8a76b072,1970GReGr...1...19K, 1971GReGr...2..321B, 1968ApJ...153L.163Z,10.1098/rspa.1974.0065,Comer:1993rx, Magli:1997qf, Gair:2001qu, Szybka:2018hoe, Mahajan:2007vw} has emerged as a particularly useful framework for quantifying the impact \cite{Lake:2006pp, Boehmer:2007az,Geralico:2012jt, Jusufi:2022jxu, Acharyya:2023rnq, Jusufi:2022jxu, Cardoso:2021wlq, Figueiredo:2023gas, Speeney:2024mas, Cardoso:2022whc, Datta:2023zmd, Shen:2023erj, Shen:2024qbb, Konoplya:2022hbl, Ovgun:2025bol}.

Generalizations to isotropic configurations with nonvanishing radial pressure have also been explored, providing a broader phenomenological landscape \cite{Datta:2023zmd}.
Most existing investigations of black holes immersed in matter distributions assume the central black hole to be non-rotating, thereby preserving exact spherical symmetry. While this assumption greatly simplifies the analysis and enables analytical progress, it is not astrophysically realistic. While these studies provide important insights, they largely assume either spherical symmetry or a non-rotating background, thereby neglecting the dynamical influence of environmental rotation on the spacetime geometry. Observational evidence strongly indicates that astrophysical black holes typically possess significant angular momentum, often approaching near-extremal values \cite{Reynolds:2019uxi, Reynolds:2020jwt}. Moreover, the host galaxies and environments in which these black holes reside are themselves expected to exhibit intrinsic rotation. Consequently, extending environmental black hole models beyond spherical symmetry and incorporating rotational degrees of freedom is essential for a realistic and comprehensive assessment of environmental effects on black hole spacetimes.

The inclusion of rotation, however, introduces substantial technical challenges. Rotation breaks spherical symmetry and generically leads to axisymmetric, stationary geometries, for which closed-form analytic solutions are rarely available. Although slowly rotating isotropic and anisotropic fluid configurations have been studied in various contexts \cite{Hartle1967,HartleThorne1968,Beltracchi:2024dfb}, embedding such matter distributions self-consistently around a rotating black hole background remains a highly nontrivial task \cite{Barausse2014,Cardoso2019}. Furthermore, algorithmic techniques based on Newman-Janis type complex transformations \cite{NewmanJanis1965,1965JMP.....6..918N, 1975JMP....16.2385G, DrakeSzekeres2000, Beltracchi:2021ris, Beltracchi:2021tcx, Kocherlakota:2024sxx, Kim:2019hfp}, which have proven remarkably successful in generating rotating vacuum or electrovac solutions, tend to fail in the presence of matter sources. In many cases, these procedures introduce unphysical features, ambiguities, or stress–energy tensors with unclear physical interpretation \cite{AzregAinou2014,HansenYunes2013}. Consequently, the solutions obtained from such algorithms lack the robustness required for reliable astrophysical modeling.

In light of these difficulties, numerical relativity approaches have become indispensable. Constructing rotating black hole solutions surrounded by anisotropic matter generally requires solving a coupled and highly nonlinear system of partial differential equations. Recently, Ref. \cite{Fernandes:2025osu} extended the Einstein-cluster framework to the case of a rotating black hole by numerically solving the full Einstein equations under some specific environmental configurations\footnote{See Ref. \cite{Cocco:2026lkr} for tidal environment.}.
While this fully numerical approach captures the complete nonlinear structure of the rotating spacetime, it remains to explore the parameter spaces and properties of the resulting geometry. The semi-analytic framework developed here offers several complementary advantages. It yields closed-form expressions that allow one to directly trace how each environmental parameter enters each observable. As it is built on a controlled perturbative expansion whose range of validity is explicit (corrections enter at order $\chi^2$), the framework is quantitatively reliable for $\chi \lesssim 0.3$. It requires only the numerical solution of a single second-order ordinary differential equation (ODE) for $\omega(r)$ rather than a coupled nonlinear partial differential equation (PDE) system, enabling rapid exploration of the multi-dimensional parameter space and the frame-dragging. The resulting orbital expressions are valid for arbitrary anisotropic matter profiles, with the Hernquist distribution serving only as a concrete astrophysical illustration.

In particular, the impact of the environment's rotation on integrals of motion has not been analyzed within a controlled semi-analytic treatment. Its effect on strong-field frequency structure has also not been explored in detail. In this work, we construct a framework and show explicitly how environmental angular velocity modifies the geometry and the associated orbital dynamics. 
This approach allows for analytic control and physical transparency while retaining the leading-order effects induced by black hole spin and frame dragging, thereby providing a complementary perspective to fully numerical treatments.

The paper is organized as follows. In Sec. \ref {sec: setup}, we introduce the theoretical framework and construct the spacetime of a slowly rotating black hole embedded in a surrounding environmental density, modeled as an anisotropic matter distribution. In Sec. \ref {sec: rotation}, we analyze rotational effects within the slow-rotation approximation and derive the differential equation governing inertial frame dragging. Orbital properties of equatorial circular timelike and null geodesics, including the conserved energy and angular momentum, as well as the location of the light ring and the innermost stable circular orbit, are investigated in Sec. \ref {sec: orbits}. In Sec. \ref {sec: results}, we present and discuss our numerical results regarding the epicyclic frequencies, highlighting the deviations from the vacuum BH behavior induced by the environment. Finally, Sec. \ref{sec:conclusion} summarizes our main findings and outlines possible observational implications.

\textit{Notations and Conventions:} Throughout this work, we employ the mostly plus metric signature. Accordingly, in $1+3$ dimensions the Minkowski metric is given by $\mathrm{diag}(-1,+1,+1,+1)$ in Cartesian coordinates. Unless stated otherwise, all calculations are performed in geometrized units, as defined previously. We express everything in the units of the BH mass $M_{\rm BH}$, and we also set $M_{\rm BH}=1$. We use $f'(r)$ to denote the first derivative of $f(r)$ with respect to $r$.

\section{Equations governing metric}\label{sec: setup}

In this work we will take a nonrotating configuration of environmental BHs and add rotation to the system perturbatively. In the context of compact stars, this has 
 been explored extensively \cite{Hartle1967,HartleThorne1968, Beltracchi:2024dfb}. In this section, we will start with such an ansatz and build the geometry. Since we are considering a spacetime describing a slowly rotating configuration, the standard slow-rotation expansion can easily be adapted. The resulting metric can be written as \cite{Hartle1967,HartleThorne1968},
\begin{align}
&ds^2 = - e^{2\nu_0(r)} \Big[ 1+2 h_0(r) + 2 h_2(r)\, P_2(\cos\theta) \Big] dt^2
\nonumber\\
& + e^{2\lambda_0(r)}\left\{1+\frac{2 e^{2\lambda_0(r)}}{r}
\Big[ m_0(r) + m_2(r)\, P_2(\cos\theta) \Big] \right\} dr^2
\nonumber\\
& + r^2 \Big[ 1+2 k_2(r)\, P_2(\cos\theta) \Big]
\Big[ d\theta^2 + \sin^2\!\theta\,\big(d\phi - \omega(r)\, dt \big)^2 \Big],
\label{kingmet1}
\end{align}
where $P_l(\cos\theta)$ denotes the Legendre polynomial of order $l$. The functions $\nu_0(r)$ and $\lambda_0(r)$ correspond to the metric components of the static, spherically symmetric background spacetime, while $h_l(r)$, $m_l(r)$, and $k_l(r)$ represent rotational corrections to the
metric. The function $\omega(r)$ arises at linear order in spin and captures the frame-dragging effect. In the present analysis, we restrict our attention to leading-order spin effects and therefore retain only terms linear in the rotation parameter. Consequently, quadratic rotational corrections associated with $h_l(r)$, $m_l(r)$, and $k_l(r)$ will be ignored in the current work.

\begin{figure*}
\centering

\subfloat[]{\includegraphics[width=0.46\textwidth]{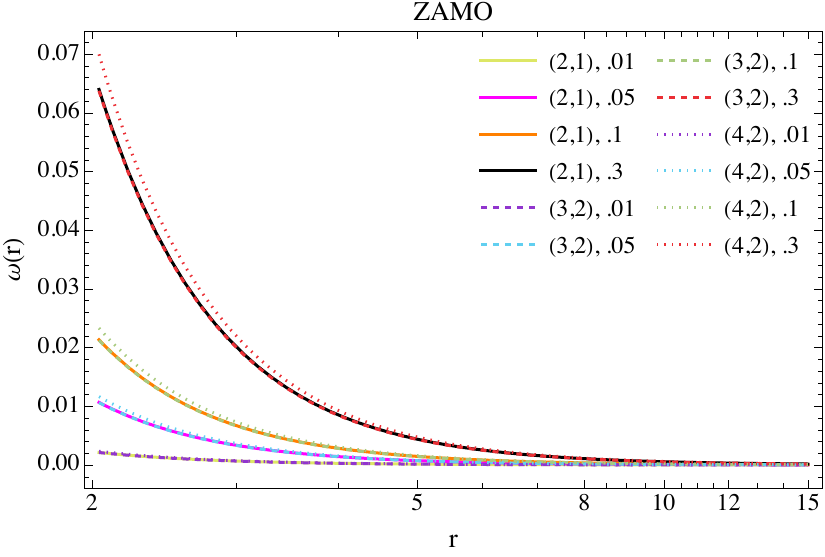}
\label{fig:omegaZAMO}}
\subfloat[]{\includegraphics[width=0.46\textwidth]{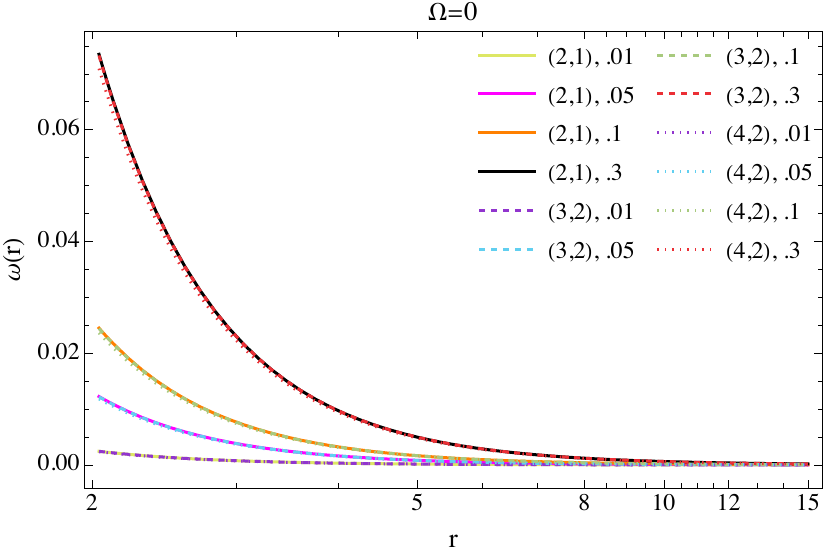}
\label{fig:omegaOmega0}}

\caption{Radial profile of $\omega(r)$ for the slowly rotating black hole surrounded by environment. Different halo configurations are labeled by $(i,j)\equiv (\log a_0,\log M_{\rm Halo})$. (a) ZAMO fluid with $\Omega=\omega(r)$. (b) Static fluid with $\Omega=0$. The resulting profile arises from the gravitational and rotational influence of the surrounding matter.}
\label{fig:omega_all}
\end{figure*}

At the background level, the spacetime is considered to be spherically symmetric and static. The geometry is described by the BH and an anisotropic fluid, whose stress-energy tensor is given by,
\begin{equation}
T^{\mu}{}_{\nu}
= \mathrm{diag}\!\left(-\rho(r),\, p_r(r),\, p_t(r),\, p_t(r)\right),
\label{stressenergy}
\end{equation}
where $\rho(r)$ denotes the energy density, $p_r(r)$ the radial pressure, and $p_t(r)$ the tangential pressure. For easy readability, we will drop the explicit demonstration of the radial functional dependence, unless specifically needed.

We assume a vanishing radial pressure, $p_r = 0$, 
which leads, through the Einstein field equations, to the relation,
\begin{equation}\label{eqn4}
p_t = \frac{m\,\rho}{2\,(\,r - 2m\,)} ,
\end{equation}
where $m(r)$ is the Misner-Sharp mass function, representing the total mass enclosed within a radius $r$. From Einstein's equations it is straightforward to compute the equations governing the background. The background metric function $\lambda_0(r)$ and $\nu_0(r)$ satisfy,
\begin{equation}\label{eqn5}
e^{-2\lambda_0} = 1 - \frac{2 m}{r},\,\, \nu_{0}' = \frac{m}{r(r-2 m)},
\end{equation}
where prime denotes radial derivative. The equation for $\lambda_0$ formally resembles the Schwarzschild exterior solution, although the mass function $m(r)$ encodes a nontrivial matter distribution. A choice of density distribution accordingly modifies $\lambda_0$ through $m(r)$. We will focus only on Hernquist-type matter distributions, leaving other profiles for future studies. We consider the density distribution to be \cite{Cardoso:2021wlq},

\begin{equation}
\label{eq: density-Hern}
    \rho= \frac{M_{\rm Halo}(a_{0}+2 M_{\rm BH})(1-2 M_{\rm BH}/r)}{2 \pi r (r+a_{0})^3}.
\end{equation}

Here, $M_{\rm BH}$ corresponds to the mass of the central black hole, $M_{\rm Halo}$ represents the mass of the halo, and $a_{0}$ is the characteristic length scale of the halo distribution within the galaxy. Different halo configuration is represented as $(i,j)\equiv ({\rm Log}\, a_0,{\rm Log}\, M_{\rm Halo})$. 
In this work we consider three configurations, $(i,j)=(2,1)$, $(3,2)$, and $(4,2)$. The $(4,2)$ case is the least compact, with compactness $\mathcal{C}\equiv M_{\rm Halo}/a_0 = 10^{-2}$, and is therefore closest to the vacuum limit. The $(2,1)$ and $(3,2)$ configurations both have $\mathcal{C}=0.1$, but differ in how close the bulk of the matter lies to the black hole. In the $(2,1)$ case the matter is located closer to the BH, whereas in $(3,2)$ it is distributed farther out.
Most studies in the literature focus primarily on the value of $\mathcal{C}$, often overlooking the importance of the closeness of matter to the black hole. Making this distinction is important, as it allows one to disentangle effects arising from the total mass content from those due to its proximity to the central object, which can lead to qualitatively different features.

These types of profiles have been extensively studied in the context of nonrotating configurations and EMRIs 
\cite{Cardoso:2022whc,Cardoso:2021wlq, Figueiredo:2023gas,Rahman:2023sof,Speeney:2024mas,Gliorio:2025cbh, Datta:2025ruh, Rahman:2025mip, Zi:2026zpw}. The primary reason behind this is the motivation that a halo distribution forms a spike near a BH \cite{Gondolo:1999ef, Sadeghian:2013laa}. Although the spike position changes in a relativistic configuration, we assume the density vanishes at the horizon. This provides a simplified and semi-analytically tractable model. We leave different configurations for future studies.

\section{Results at linear order in spin} \label{sec: rotation}

We assume a slowly rotating black hole-matter configuration. The matter surrounding the black hole can be modeled by a generic anisotropic fluid with an associated stress–energy tensor as,
\begin{equation}\label{eq:envT}
     T_{\mu\nu} = \Bar{\rho} u_{\mu}u_{\nu} + \bar{p}_r k_{\mu}k_{\nu} + \bar{p}_t \Pi_{\mu\nu}\ ,
\end{equation}
where we call $\bar{p}_r$ and $\bar{p}_t$ as radial and tangential pressures, $u^\mu$ is the fluid four velocity and $k^\mu$ is a unit space-like radial vector orthogonal to the later, such that $-u_{\mu}u^{\mu} = k_{\mu}k^{\mu}=1$ and $u_{\mu}k^{\mu}=0$ \cite{Bowers:1974tgi,Doneva:2012rd,Raposo:2018rjn}. 
The projector on the surface orthogonal to the 4-velocity and $k^\mu$ 
is given by $\Pi_{\mu\nu}= g_{\mu\nu} + u_{\mu}u_{\nu} - k_{\mu}k_{\nu}$, with $u^\mu \Pi_{\mu\nu}X^\nu=k^\mu \Pi_{\mu\nu}X^\nu=0$, for a generic vector $X^\nu$. The four-velocity of the fluid is chosen as,
\begin{align}
u^t &= \left(-g_{tt} - 2 \Omega\, g_{t\phi}
- \Omega^2 g_{\phi\phi}\right)^{-1/2}, \nonumber\\
u^\phi &= \Omega\, u^t, \qquad
u^r = u^\theta = 0,
\label{4vhartle}
\end{align}
where $\Omega \equiv d\phi/dt$ is the angular velocity measured by an observer at infinity. Note, $\Omega$ remains unconstrained at this stage. Later we will choose a different functional form of $\Omega$ to study its impact on the geometry, and as a consequence, on the observable quantities.

Having specified the governing equations for the background spacetime and the anisotropic matter content, we now investigate the effects induced by rotation. We work within the slow-rotation approximation and retain only terms linear in spin. We leave the higher-order spin effects for future studies. Within this approximation, the only non-vanishing rotational correction to the metric arises through the off-diagonal component $g_{t\phi} = - r^2 \sin^2\theta\, \omega(r)$, which encodes the dragging of inertial frames.
For a generic anisotropic matter distribution described by the stress-energy tensor in Eq. \eqref{stressenergy}, the linearized Einstein equations yield a single nontrivial equation governing the rotational perturbation. We find the equation to be: 
\begin{align}
\label{newomega}
 0=& \, \omega''+\left(4 - \frac{4\pi r^3 \rho}{r - 2m}\right)\frac{\omega'}{r} \nonumber \\
 &+ \frac{8\pi r (3m - 2r)\rho\,\omega }{ (r - 2m)^2}
+ \frac{16\pi r (\rho+p_t)\Omega}{r-2m}.
\end{align}
Eq. \eqref{newomega} extends the Hartle-Thorne slow-rotation equation to the case of anisotropic matter. Now using the background Einstein equations for the static anisotropic fluid Eqns. \eqref{eqn4} and \eqref{eqn5}, Eq. \eqref{newomega} can be recast into a second-order differential equation directly for the frame-dragging function $\omega(r)$ as,
\begin{align}
\left(4 - \frac{4\pi r^3 \rho}{r - 2m}\right)\frac{\omega'}{r}
+ \frac{8\pi r (3m - 2r)\rho\,(\omega - \Omega)}{ (r - 2m)^2}
+ \omega'' = 0 .
\label{newomega1}
\end{align}

This equation governs the radial behavior of inertial-frame dragging in the presence of anisotropic matter and explicitly demonstrates how deviations from pressure isotropy modify the rotational response of the spacetime. The appearance of the combination $(\omega-\Omega)$ explicitly shows that inertial frame dragging is sourced not only by the black hole spin but also by the relative rotation between spacetime and the surrounding fluid. In the vacuum limit, $\rho \rightarrow 0$, Eq.~\eqref{newomega1} reduces to the standard frame-dragging equation for the slowly rotating vacuum geometry. Importantly, Eq. (\ref{newomega1}) is applicable to any density profiles. The subsequent choice of a Hernquist-type distribution serves only to provide a concrete astrophysical illustration. All qualitative features discussed below follow from the structure of the rotational equation itself and are not tied to this specific profile.

\begin{table}[ht]
\centering
\begin{tabular}{ccccc}
\hline
$\Omega$ &  Profile & $\chi (10^6)$ & $\chi (2.1)$ \\
\hline
\text{ZAMO} & (2, 1) & 0.01 & 0.009 \\
 \text{ZAMO} & (2, 1) & 0.05 & 0.0453 \\
 \text{ZAMO} & (2, 1) & 0.1 & 0.0905 \\
 \text{ZAMO} & (2, 1) & 0.3 & 0.2716 \\
 \text{ZAMO} & (3, 2) & 0.01 & 0.009 \\
 \text{ZAMO} & (3, 2) & 0.05 & 0.0451 \\
 \text{ZAMO} & (3, 2) & 0.1 & 0.0902 \\
 \text{ZAMO} & (3, 2) & 0.3 & 0.2706 \\
 \text{ZAMO} & (4, 2) & 0.01 & 0.0099 \\
 \text{ZAMO} & (4, 2) & 0.05 & 0.0495 \\
 \text{ZAMO} & (4, 2) & 0.1 & 0.099 \\
 \text{ZAMO} & (4, 2) & 0.3 & 0.2969 \\
0  &  (2, 1) & 0.01 & 0.0104 \\
 0  &  (2, 1) & 0.05 & 0.0519 \\
 0  &  (2, 1) & 0.1 & 0.1039 \\
 0  &  (2, 1) & 0.3 & 0.3117 \\
 0  &  (3, 2) & 0.01 & 0.0104 \\
 0  &  (3, 2) & 0.05 & 0.0518 \\
 0  &  (3, 2) & 0.1 & 0.1036 \\
 0  &  (3, 2) & 0.3 & 0.3109 \\
 0  &  (4, 2) & 0.01 & 0.01 \\
 0  &  (4, 2) & 0.05 & 0.0501 \\
 0  &  (4, 2) & 0.1 & 0.1003 \\
 0  &  (4, 2) & 0.3 & 0.301 \\
\hline
\end{tabular}
\caption{In the table we report the spin values near the horizon for a chosen spin value at large distance $(r=10^6)$. Different halo configuration is represented as $(i,j)\equiv (Loga_0,Log M_{\rm Halo})$. We define $\chi \equiv \omega(r)r^3/2$. $\chi(r)$ should be interpreted as an effective local spin parameter, not a conserved quantity. We fix the asymptotic spin and report the corresponding local value $\chi(r)$. Spin value clearly depends on the halo configuration and the choice of $\Omega$. }
\label{tab: spin evolution}
\end{table}

Using the density profile in Eq. (\ref{eq: density-Hern}) first we numerically solve for metric components from Eq. (\ref{eqn5}). The mass function directly provides a solution for $\lambda_0$. To compute $\nu_0$ we solve the differential equation with $\nu_0(10^6)=0$. Then we use these results in Eq. (\ref{newomega1}) to compute $\omega$. As a boundary condition we choose $\omega(r=10^6) = 2\chi/(10^6)^3$ and $\omega'(r=10^6) = -6\chi/(10^6)^4$. In the limit $\rho \to 0$, these boundary conditions reproduce the standard slow-rotation vacuum solution. We find solutions for different choices of $\chi$. For the fluid angular velocity $\Omega$ we choose two specific configurations. Firstly, $\Omega=\omega$, this corresponds to a fluid corotating with the local zero angular momentum observers, represented as ZAMO. For the second configuration we choose the static observers, namely, $\Omega=0$. The results are provided in Fig. \ref{fig:omegaZAMO} and \ref{fig:omegaOmega0}. The choices $\Omega=\omega(r)$ and $\Omega=0$ bracket physically distinct limits of the fluid’s angular velocity and allow us to isolate the impact of matter rotation on frame dragging and orbital dynamics. More realistic rotation laws can be incorporated within the same framework and are expected to interpolate between the limiting behaviors presented here.

Different halo configuration is represented as $(i,j)\equiv (Loga_0,Log M_{\rm Halo})$. For a given halo configuration we choose the asymptotic spin to be $0.01,\, 0.05,\, 0.1,\, 0.3$. In the left panel we demonstrate ZAMO configuration while in the right $\Omega=0$. We find a clear dependence on the choice of $\Omega$ in the plots. As expected, for larger spins we find the magnitude of $\omega$ to be larger. The results indicate that the configuration is more sensitive to the ratio $a_0/M_{\rm Halo}$ than to the individual values of $ a_0$ and $ M _ {\rm Halo} $. This can be seen by comparing (2, 1) and (3, 2) results, which seem to be very close to each other in all of the plots. At the same time the (4, 2) configuration seems to be more different. We also notice that the choice of $\Omega$ determines the sensitivity of $\omega$ on the profile configuration. This can be seen by comparing the left and the right panel. In the ZAMO configuration the different profiles are more separated from each other compared to the static configurations. Therefore, the choice of $\Omega$ has a discernible impact on the metric. This aspect needs to be investigated further as it has the potential to provide us with information regarding the fluid flows near the BHs. 

In Table  \ref{tab: spin evolution} we provide the spin values defined as $\chi(r) \equiv \omega(r) r^3/2$ at $r=2.1$. In the vacuum limit, this definition reduces to the standard slow-rotation Kerr parameter, while in the presence of matter it should be interpreted as an effective local rotational measure. As expected, increasing asymptotic spin leads to larger spin near the horizon. 
For the ZAMO configuration, the near-horizon spin is systematically smaller than its asymptotic value for all the halo profiles, whereas for the static configuration, it is consistently larger. In both cases, the deviation remains small in the slow-rotation regime, although larger differences may appear in fully nonlinear high-spin solutions. Table \ref{tab: spin evolution} also shows a clear dependence on halo profile and on the choice of $\Omega$. 
These results indicate that fluid motion near the black hole can modify the spacetime geometry and may be relevant for EMRIs, which are sensitive to near-BH structure. Environmental rotation therefore induces frame dragging in a profile-dependent manner, which directly propagates into the orbital observables studied in the following section.

\section{Equatorial Circular Geodesics}\label{sec: orbits}

Having established the structure of the slowly rotating spacetime and determined the frame-dragging function $\omega(r)$ from the rotational field equations in hand, we now investigate the motion of test particles in this geometry. We restrict our analysis to equatorial and circular timelike geodesics, which are directly relevant to astrophysical observables such as orbital frequencies, accretion dynamics, the location of characteristic orbits, and the epicyclic frequencies.

Owing to the stationarity and axial symmetry of the spacetime, particle motion admits two conserved quantities associated with the Killing vectors $\partial_t$ and $\partial_\phi$. The orbital angular velocity, $\Omega_o$, as measured by an observer at infinity, is defined as,
\begin{equation}
\Omega_o \equiv \frac{d\phi}{dt} = \frac{u^{\phi}}{u^{t}},
\end{equation}
where $u^\mu$ denotes the four-velocity of the particle. The normalization condition for timelike trajectories $u^\mu u_\mu = -1$ yields the time component of the four-velocity as,
\begin{equation}
u^t =
\frac{1}{\sqrt{-\left(g_{tt} + 2\Omega_o g_{t\phi} + \Omega_o^2 g_{\phi\phi}\right)}} .
\end{equation}

The conserved energy $E$ and angular momentum $L$ per unit rest mass are defined as,
\begin{align}
\label{eq: exact E and L}
E &\equiv -u_t
= -\left(g_{tt} u^t + g_{t\phi} u^\phi\right)
= -u^t\left(g_{tt} + \Omega_o g_{t\phi}\right),\nonumber \\
L &\equiv u_\phi
= g_{t\phi} u^t + g_{\phi\phi} u^\phi
= u^t\left(g_{t\phi} + \Omega_o g_{\phi\phi}\right),
\end{align}
which remain constant along the geodesic as a consequence of the underlying symmetries. The values of the energy and angular momentum can be computed directly once the orbital frequencies are computed, as discussed in the next section.

\begin{table}[ht]
\centering
\begin{tabular}{ccccc}
\hline
$\Omega$ & Orbit & Profile & $\chi$ & $r_{LR}$ \\
\hline 
 0 & \text{Prograde} & (2,1) & 0.01 & 2.989 \\
 0 & \text{Prograde} & (2,1) & 0.05 & 2.934 \\
 0 & \text{Prograde} & (2,1) & 0.1 & 2.860 \\
 0 & \text{Prograde} & (2,1) & 0.3 & 2.466 \\
 0 & \text{Prograde} & (3,2) & 0.01 & 2.987 \\
 0 & \text{Prograde} & (3,2) & 0.05 & 2.932 \\
 0 & \text{Prograde} & (3,2) & 0.1 & 2.858 \\
 0 & \text{Prograde} & (3,2) & 0.3 & 2.466 \\
 0 & \text{Prograde} & (4,2) & 0.01 & 2.988 \\
 0 & \text{Prograde} & (4,2) & 0.05 & 2.940 \\
 0 & \text{Prograde} & (4,2) & 0.1 & 2.875 \\
 0 & \text{Prograde} & (4,2) & 0.3 & 2.553 \\
 0 & \text{Retrograde} & (2,1) & 0.01 & 3.016 \\
 0 & \text{Retrograde} & (2,1) & 0.05 & 3.067 \\
 0 & \text{Retrograde} & (2,1) & 0.1 & 3.128 \\
 0 & \text{Retrograde} & (2,1) & 0.3 & 3.343 \\
 0 & \text{Retrograde} & (3,2) & 0.01 & 3.013 \\
 0 & \text{Retrograde} & (3,2) & 0.05 & 3.065 \\
 0 & \text{Retrograde} & (3,2) & 0.1 & 3.125 \\
 0 & \text{Retrograde} & (3,2) & 0.3 & 3.339 \\
 0 & \text{Retrograde} & (4,2) & 0.01 & 3.012 \\
 0 & \text{Retrograde} & (4,2) & 0.05 & 3.057 \\
 0 & \text{Retrograde} & (4,2) & 0.1 & 3.111 \\
 0 & \text{Retrograde} & (4,2) & 0.3 & 3.304 \\
\hline
\end{tabular}
\caption{Prograde and retrograde light-ring radii $r_{LR}$ for a static fluid ($\Omega=0$), considering different halo configurations and spin values.}
\label{tab:Light-ring radii Omega 0}
\end{table}

\subsection{Orbital Frequency of circular orbits}

The condition for circular equatorial motion follows from extremizing the radial effective potential, leading to,
\begin{equation}
\Omega_o \, \partial_r g_{\phi\phi}
= - \partial_r g_{t\phi}
\pm \sqrt{\left(\partial_r g_{t\phi}\right)^2
- \partial_r g_{tt}\, \partial_r g_{\phi\phi}}~,
\end{equation}
where the upper (lower) sign corresponds to co-rotating (counter-rotating) orbits with respect to the spin of the compact object.

Substituting the slowly rotating metric in Eq.~\eqref{kingmet1} and working consistently to linear order in the spin parameter, only terms proportional to the frame-dragging function $\omega(r)$ are retained. The orbital angular frequency of equatorial circular orbits is then given by,
\begin{equation}
\Omega_{\pm}
= \pm e^{\nu_0}\sqrt{\frac{\,\nu_0'\,}{r}}
+ \omega + \frac{1}{2} r\,\omega',
\label{Omega_pm}
\end{equation}
where a prime denotes differentiation with respect to the radial coordinate $r$. The first term gives the circular-orbit frequency in the nonrotating background, while the remaining terms capture the leading frame-dragging corrections. These expressions determine the locations of the light ring and the innermost stable circular orbit discussed in the next section.

\begin{table}[ht]
\centering
\begin{tabular}{ccccc}
\hline
$\Omega$ & Orbit & Profile & $\chi$ & $r_{LR}$ \\
\hline
\text{ZAMO} & \text{Prograde} & (2,1) & 0.01 & 2.991 \\
 \text{ZAMO} & \text{Prograde} & (2,1) & 0.05 & 2.943 \\
 \text{ZAMO} & \text{Prograde} & (2,1) & 0.1 & 2.879 \\
 \text{ZAMO} & \text{Prograde} & (2,1) & 0.3 & 2.562 \\
 \text{ZAMO} & \text{Prograde} & (3,2) & 0.01 & 2.989 \\
 \text{ZAMO} & \text{Prograde} & (3,2) & 0.05 & 2.941 \\
 \text{ZAMO} & \text{Prograde} & (3,2) & 0.1 & 2.877 \\
 \text{ZAMO} & \text{Prograde} & (3,2) & 0.3 & 2.561 \\
 \text{ZAMO} & \text{Prograde} & (4,2) & 0.01 & 2.988 \\
 \text{ZAMO} & \text{Prograde} & (4,2) & 0.05 & 2.941 \\
 \text{ZAMO} & \text{Prograde} & (4,2) & 0.1 & 2.877 \\
 \text{ZAMO} & \text{Prograde} & (4,2) & 0.3 & 2.561 \\
 \text{ZAMO} & \text{Retrograde} & (2,1) & 0.01 & 3.014 \\
 \text{ZAMO} & \text{Retrograde} & (2,1) & 0.05 & 3.059 \\
 \text{ZAMO} & \text{Retrograde} & (2,1) & 0.1 & 3.113 \\
 \text{ZAMO} & \text{Retrograde} & (2,1) & 0.3 & 3.305 \\
 \text{ZAMO} & \text{Retrograde} & (3,2) & 0.01 & 3.012 \\
 \text{ZAMO} & \text{Retrograde} & (3,2) & 0.05 & 3.057 \\
 \text{ZAMO} & \text{Retrograde} & (3,2) & 0.1 & 3.110 \\
 \text{ZAMO} & \text{Retrograde} & (3,2) & 0.3 & 3.301 \\
 \text{ZAMO} & \text{Retrograde} & (4,2) & 0.01 & 3.011 \\
 \text{ZAMO} & \text{Retrograde} & (4,2) & 0.05 & 3.056 \\
 \text{ZAMO} & \text{Retrograde} & (4,2) & 0.1 & 3.109 \\
 \text{ZAMO} & \text{Retrograde} & (4,2) & 0.3 & 3.300 \\
\hline
\end{tabular}
\caption{Prograde and retrograde light-ring radii $r_{LR}$ for a ZAMO fluid configuration, considering different halo configurations and spin values.}
\label{tab:Light-ring radii ZAMO} 
\end{table}

\subsection{Light Ring and ISCO}

One of the key features of BH space times is the existence of circular orbits for photons, namely, the light ring. The location of the light ring (circular photon orbit) in the equatorial plane is determined by imposing the null-geodesic condition together with the
extremization of the effective potential. This leads to the following equation for the light-ring radius as,
\begin{equation}
\label{eq:LR}
e^{\nu_0} \left( r \nu_0' - 1 \right)
\pm r^2 \sqrt{ r \, \nu_0' } \, \omega' = 0 ,
\end{equation}
where the upper (lower) sign corresponds to prograde (retrograde) light rings. This expression explicitly demonstrates how frame dragging shifts the photon sphere relative to its nonrotating location. The above result is consistent with the slow-spin Hartle-Thorne result found in Ref. \cite{Abramowicz:2003rc}.

\begin{table}[ht]
\centering
\begin{tabular}{cccccc}
\hline
 $\Omega$ & Orbit & Profile & $\chi$ & $r_{ISCO}$ & $\Omega_{ISCO}$ \\
\hline
 \text{ZAMO} & \text{Prograde} & (2,1) & 0.01 & 5.825 & 0.065 \\
 \text{ZAMO} & \text{Prograde} & (2,1) & 0.05 & 5.700 & 0.067 \\
 \text{ZAMO} & \text{Prograde} & (2,1) & 0.1 & 5.536 & 0.070 \\
 \text{ZAMO} & \text{Prograde} & (2,1) & 0.3 & 4.761 & 0.085 \\
 \text{ZAMO} & \text{Prograde} & (3,2) & 0.01 & 5.949 & 0.062 \\
 \text{ZAMO} & \text{Prograde} & (3,2) & 0.05 & 5.816 & 0.064 \\
 \text{ZAMO} & \text{Prograde} & (3,2) & 0.1 & 5.643 & 0.067 \\
 \text{ZAMO} & \text{Prograde} & (3,2) & 0.3 & 4.828 & 0.083 \\
 \text{ZAMO} & \text{Prograde} & (4,2) & 0.01 & 5.967 & 0.068 \\
 \text{ZAMO} & \text{Prograde} & (4,2) & 0.05 & 5.833 & 0.070 \\
 \text{ZAMO} & \text{Prograde} & (4,2) & 0.1 & 5.658 & 0.073 \\
 \text{ZAMO} & \text{Prograde} & (4,2) & 0.3 & 4.838 & 0.090 \\
 \text{ZAMO} & \text{Retrograde} & (2,1) & 0.01 & 5.885 & -0.064 \\
 \text{ZAMO} & \text{Retrograde} & (2,1) & 0.05 & 6.004 & -0.063 \\
 \text{ZAMO} & \text{Retrograde} & (2,1) & 0.1 & 6.146 & -0.061 \\
 \text{ZAMO} & \text{Retrograde} & (2,1) & 0.3 & 6.669 & -0.054 \\
 \text{ZAMO} & \text{Retrograde} & (3,2) & 0.01 & 6.014 & -0.061 \\
 \text{ZAMO} & \text{Retrograde} & (3,2) & 0.05 & 6.140 & -0.060 \\
 \text{ZAMO} & \text{Retrograde} & (3,2) & 0.1 & 6.292 & -0.058 \\
 \text{ZAMO} & \text{Retrograde} & (3,2) & 0.3 & 6.859 & -0.051 \\
 \text{ZAMO} & \text{Retrograde} & (4,2) & 0.01 & 6.032 & -0.067 \\
 \text{ZAMO} & \text{Retrograde} & (4,2) & 0.05 & 6.160 & -0.065 \\
 \text{ZAMO} & \text{Retrograde} & (4,2) & 0.1 & 6.314 & -0.063 \\
 \text{ZAMO} & \text{Retrograde} & (4,2) & 0.3 & 6.887 & -0.056 \\
\hline
\end{tabular}
\caption{ISCO radius and orbital frequencies for the ZAMO configuration for prograde and retrograde orbits for the different halo configurations and spin values. The Schwarzschild limit corresponds to $r_{\rm ISCO}=6$}
\label{tab:ISCO ZAMO}
\end{table}

Using Eq. (\ref{eq:LR}), we computed the light-ring (LR) radii $(r_{LR})$ for massless particles. The results are provided in Table \ref{tab:Light-ring radii Omega 0} and \ref{tab:Light-ring radii ZAMO} for the static and the ZAMO case, respectively. For prograde motion, the light-ring radius decreases with increasing $\chi$, indicating that halo rotation shifts the prograde light rings slightly inward. Conversely, retrograde light rings move outward as $\chi$ increases, reflecting the opposing effect of the halo’s rotation on counter-rotating photons. Light ring position also depends on the profile parameters represented by $(i,j)$ for both prograde and retrograde orbits.
Comparing the two rotation prescriptions, the differences between ZAMO and $\Omega=0$ configuration increase with increasing halo compactness and spin, with the larger deviations occurring at higher $\chi$ for both the prograde and retrograde orbits. Overall, these results demonstrate that halo rotation and configuration have a noticeable but moderate effect on light-ring positions. It remains to be seen how higher spin corrections affect these results.


Another important orbit near BHs is the innermost stable circular orbit (ISCO). It is the smallest radius at which matter can orbit a black hole in a stable circular path. We find that for ISCO, the orbits must satisfy,

\begin{align}
\label{eq:ISCO condition}
  &r^3
   (r-2 m) m \omega '' +r^2 (3 r-10 m) m \omega '\nonumber\\
   &\pm \sqrt{\frac{m}{r-2 m}} e^{\nu_0} \left(r^2 m'+(r-6 m) m\right) =0.
\end{align}
Eq. (\ref{eq:ISCO condition}) reproduces the correct ISCO position found in Ref. \cite{Shibata:1998xw, Abramowicz:2003rc} for vacuum.

\begin{table}[ht]
\centering
\begin{tabular}{cccccc}
\hline
$\Omega$ & Orbit & Profile & $\chi$ & $r_{ISCO}$ & $\Omega_{ISCO}$ \\
\hline
 0 & \text{Prograde} & (2,1) & 0.01 & 5.820 & 0.065 \\
 0 & \text{Prograde} & (2,1) & 0.05 & 5.677 & 0.067 \\
 0 & \text{Prograde} & (2,1) & 0.1 & 5.485 & 0.071 \\
 0 & \text{Prograde} & (2,1) & 0.3 & 4.557 & 0.090 \\
 0 & \text{Prograde} & (3,2) & 0.01 & 5.944 & 0.062 \\
 0 & \text{Prograde} & (3,2) & 0.05 & 5.791 & 0.065 \\
 0 & \text{Prograde} & (3,2) & 0.1 & 5.588 & 0.068 \\
 0 & \text{Prograde} & (3,2) & 0.3 & 4.614 & 0.088 \\
 0 & \text{Prograde} & (4,2) & 0.01 & 5.967 & 0.068 \\
 0 & \text{Prograde} & (4,2) & 0.05 & 5.831 & 0.070 \\
 0 & \text{Prograde} & (4,2) & 0.1 & 5.653 & 0.073 \\
 0 & \text{Prograde} & (4,2) & 0.3 & 4.820 & 0.091 \\
 0 & \text{Retrograde} & (2,1) & 0.01 & 5.890 & -0.064 \\
 0 & \text{Retrograde} & (2,1) & 0.05 & 6.025 & -0.062 \\
 0 & \text{Retrograde} & (2,1) & 0.1 & 6.187 & -0.060 \\
 0 & \text{Retrograde} & (2,1) & 0.3 & 6.775 & -0.053 \\
 0 & \text{Retrograde} & (3,2) & 0.01 & 6.018 & -0.061 \\
 0 & \text{Retrograde} & (3,2) & 0.05 & 6.163 & -0.059 \\
 0 & \text{Retrograde} & (3,2) & 0.1 & 6.337 & -0.057 \\
 0 & \text{Retrograde} & (3,2) & 0.3 & 6.976 & -0.050 \\
 0 & \text{Retrograde} & (4,2) & 0.01 & 6.033 & -0.067 \\
 0 & \text{Retrograde} & (4,2) & 0.05 & 6.162 & -0.065 \\
 0 & \text{Retrograde} & (4,2) & 0.1 & 6.318 & -0.063 \\
 0 & \text{Retrograde} & (4,2) & 0.3 & 6.897 & -0.056 \\
\hline
\end{tabular}
\caption{Prograde and retrograde ISCO radius and orbital frequencies for $\Omega = 0$, considering different halo configurations and spin values. The Schwarzschild limit corresponds to $r_{\rm ISCO}=6$}
\label{tab:ISCO $0$}
\end{table}

We use Eq. (\ref{eq:ISCO condition}) to compute the ISCO radii numerically. The results are reported in Tables \ref{tab:ISCO ZAMO} and \ref{tab:ISCO $0$} for the ZAMO and $\Omega=0$ configurations, respectively, with the last column giving the ISCO orbital frequency $\Omega_{\rm ISCO}$. In the Schwarzschild limit, $r_{\rm ISCO}=6$ and $\Omega_{\rm ISCO}\simeq 0.068$, in agreement with Refs.~\cite{Shibata:1998xw, Abramowicz:2001bi}.
We consider halo profiles $(i,j)$ and rotation parameters $\chi$. For prograde motion, the ISCO radius decreases with increasing $\chi$, while for retrograde motion it increases, reflecting the competing effect of halo rotation. The ZAMO prescription typically yields slightly larger prograde ISCO radii than the $\Omega=0$ case, while smaller ones for retrograde orbits.
Overall, halo rotation and profile parameters modify the ISCO structure, with $\chi$ driving the dominant branching between prograde and retrograde orbits.

\subsection{Energy and Angular Momentum}

Having established the geometric properties, we now quantify deviations of the orbital energy and angular momentum from their vacuum values. This is particularly relevant for EMRIs, where waveform phasing depends directly on conservative orbital quantities. Using Eq. (\ref{eq: exact E and L}), we compute the conserved energy and angular momentum per unit mass as,
\begin{equation}
2\sqrt{\left(1 - r\nu_0'\right)}\, E_{\pm}
= 2e^{\nu_0}
\pm
\frac{r\,\sqrt{r\nu_0'}
\left[2\omega\left(r\nu_0' - 1\right) - r\omega'\right]}
{r\nu_0' - 1},
\end{equation}
where, $(+)$ and $(-)$ respectively represent the prograde and retrograde orbits. This explicitly illustrates how frame dragging modifies the energetics of equatorial circular motion at linear order in the spin.

In an analogous manner, the conserved angular momentum per unit mass associated with equatorial circular motion is given by,
\begin{equation}
\sqrt{ \left(1-r \nu_0'\right)} \, L_{\pm}
= \frac{r^3 \omega '}{2 e^{ \nu_0}(1-r \nu_0')}
\pm r\sqrt{r  \nu_0'}
,
\end{equation}
where the upper (lower) sign corresponds to prograde (retrograde) orbits. The above energy and angular momentum expressions reproduce the Hartle-Thorne expressions found in Ref. \cite{Abramowicz:2003rc}.

\begin{figure*}
\centering

\subfloat[]{\includegraphics[width=0.46\textwidth]{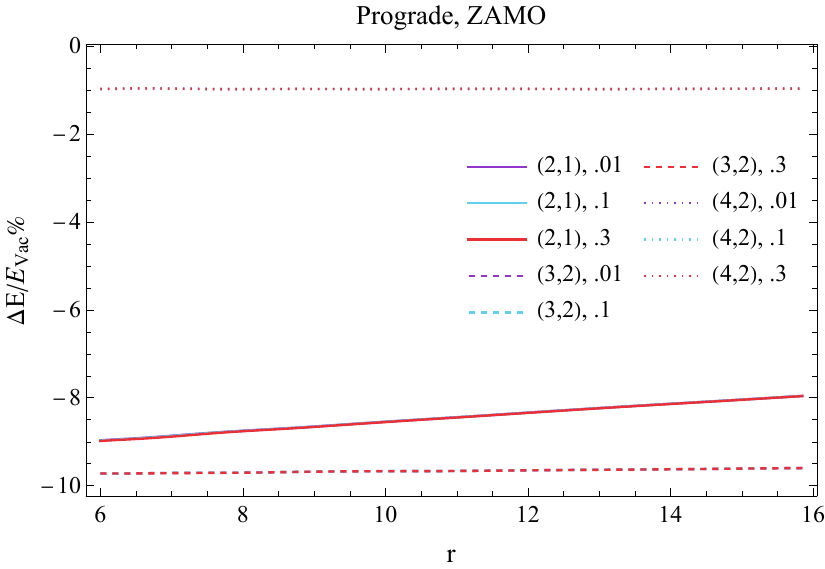}
\label{fig:EZAMO}}
\hfill
\subfloat[]{\includegraphics[width=0.46\textwidth]{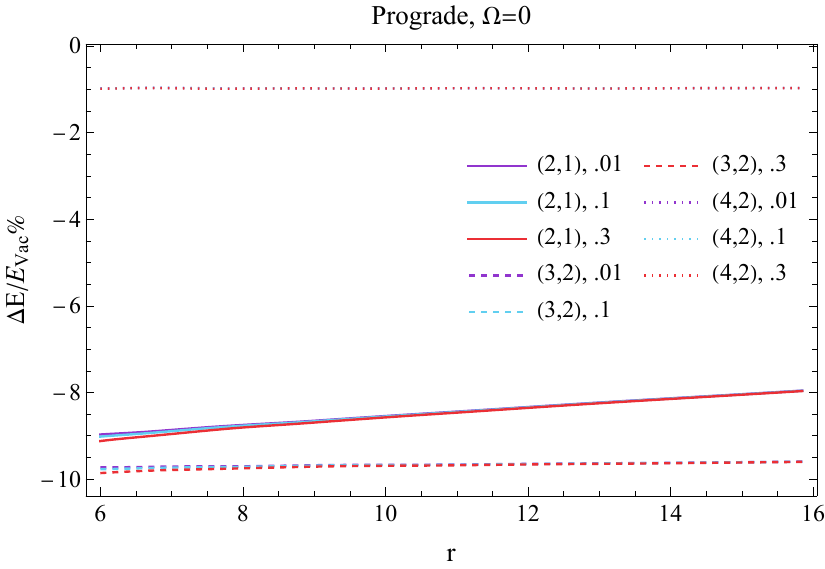}
\label{fig:EOmega0}}

\subfloat[]{\includegraphics[width=0.46\textwidth]{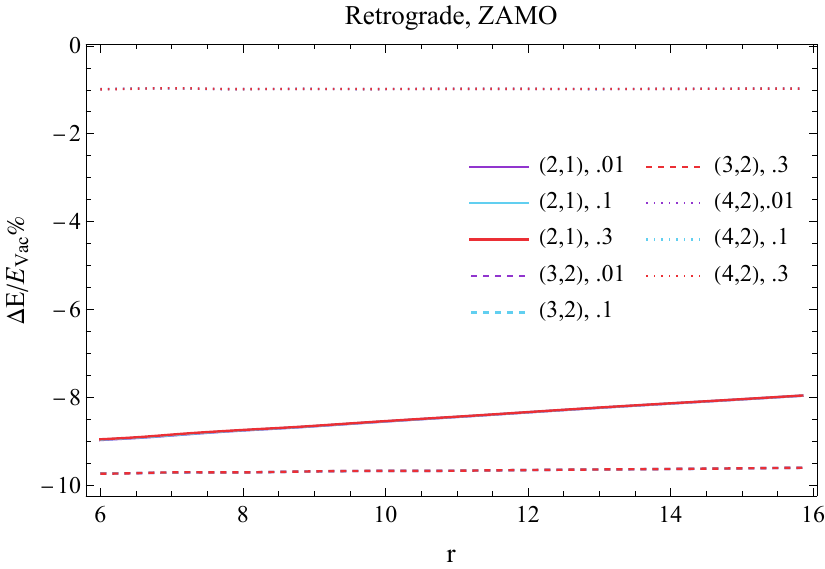}
\label{fig:ERet-ZAMO}}
\hfill
\subfloat[]{\includegraphics[width=0.46\textwidth]{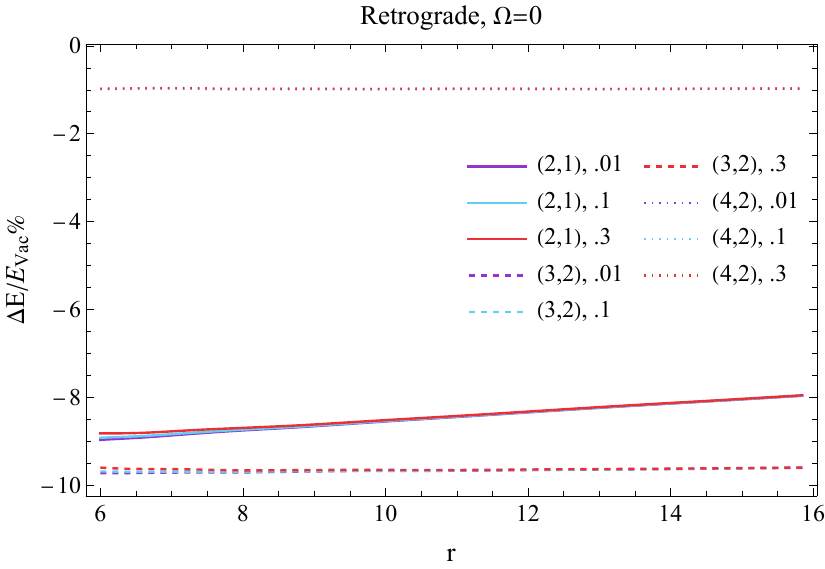}
\label{fig:ERet-Omega0}}

\caption{Relative fractional deviation of the conserved energy per unit rest mass, $\Delta E/E_{\rm Vac}\%$, for equatorial circular timelike orbits compared to the slow-rotation vacuum case. Upper (lower) panels correspond to prograde (retrograde) motion. Left panels show the ZAMO prescription, right panels the static fluid ($\Omega=0$). Different curves denote halo configurations. Deviations arise from the combined effects of halo gravity and modified frame dragging. All curves are retained to show the systematic trend that differences between ZAMO and static fluid grow with increasing spin, becoming noticeable even at $\chi=0.3$ and in near BH, which is the EMRI relevant region (see right panel).}

\label{fig:E change}
\end{figure*}

\begin{figure*}
\centering

\subfloat[]{\includegraphics[width=0.46\textwidth]{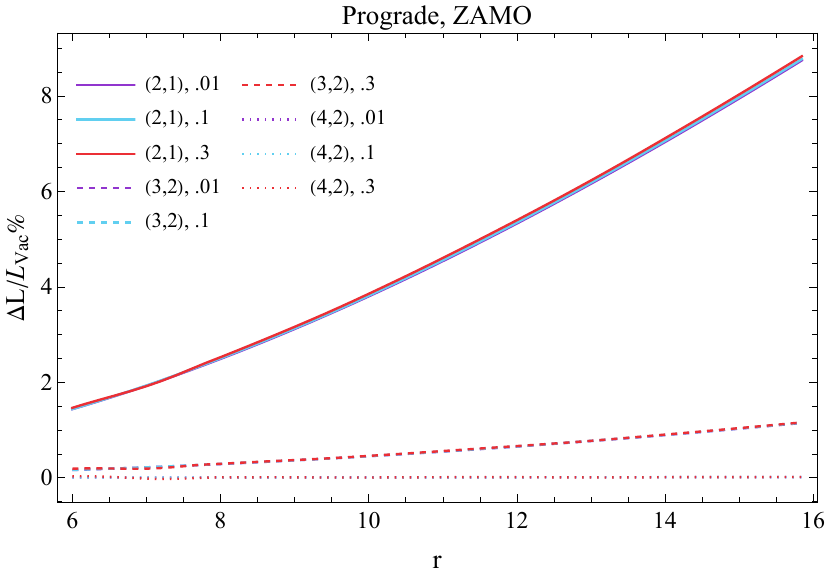}
\label{fig:LZAMO}}
\hfill
\subfloat[]{\includegraphics[width=0.46\textwidth]{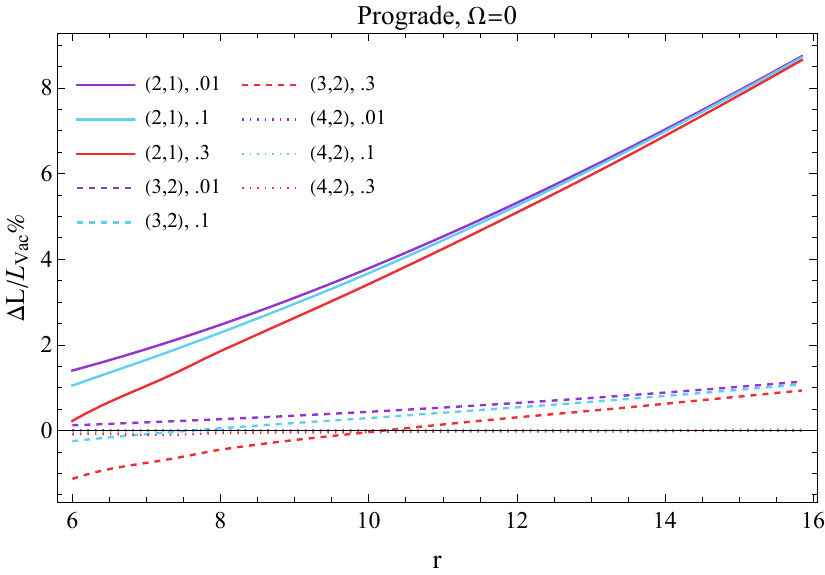}
\label{fig:LOmega0}}

\subfloat[]{\includegraphics[width=0.46\textwidth]{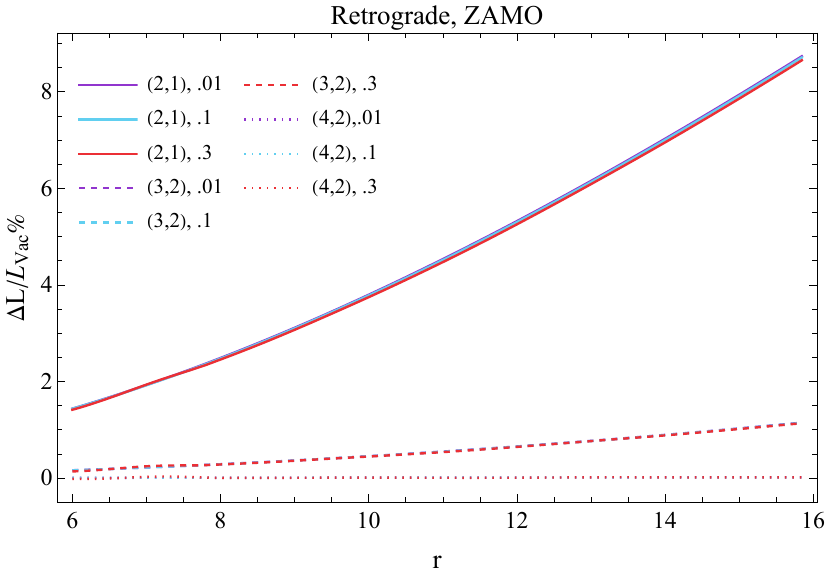}
\label{fig:LRet-ZAMO}}
\hfill
\subfloat[]{\includegraphics[width=0.46\textwidth]{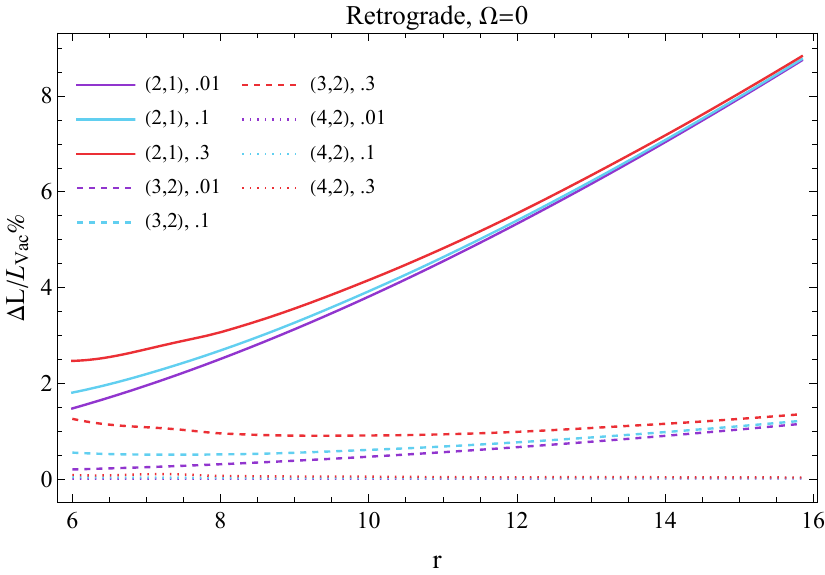}
\label{fig:LRet-Omega0}}

\caption{Relative fractional deviation of the conserved angular momentum per unit rest mass, $\Delta L/L_{\rm Vac}\%$, for equatorial circular timelike orbits compared to the slow-rotation vacuum case. Upper (lower) panels correspond to prograde $(+)$ (retrograde $(-)$) motion. Left panels show the ZAMO prescription with $\Omega=\omega(r)$, while right panels correspond to a static fluid with $\Omega=0$. Different curves represent distinct halo configurations. The radial dependence and prograde–retrograde asymmetry arise from the combined influence of halo gravity and environmental frame dragging. The differences between static and ZAMO fluid grow with increasing spin, becoming most prominent at $\chi=0.3$ and near the BH. In the right panel, individual halo profiles are visibly split into different curves, unlike in the left panel. It demonstrates that the shift in the conserved quantities is dependent on the velocity properties of the fluid.}
\label{fig:L change}
\end{figure*}

Fig. \ref{fig:E change} and \ref{fig:L change} show the relative percentage deviations of the conserved energy and angular momentum per unit rest mass,
$\Delta E/E_{\rm Vac}\%$ and $\Delta L/L_{\rm Vac}\%$, for equatorial circular timelike orbits in a slowly rotating spacetime immersed in an environment, compared to the slow rotation limit of the vacuum case, where $\Delta E=E_{\pm} - E_{\rm Vac}$ and $\Delta L=L_{\pm} - L_{\rm Vac}$, with ``${\rm Vac}$" used as a shorthand for vacuum. The results are presented for both prograde $(+)$ and retrograde $(-)$ orbits, and for both the ZAMO and the static configuration. The plots of the energy and angular momentum themselves are provided in Fig. \ref{fig:E-L all} in \ref{app:Energy and angular momentum}.

The energy deviation, $\Delta E/E_{\rm Vac}\%$, exhibits a mild independence from the radial coordinate $r$ in the shown region except for the $(2,1)$ configuration. 
The magnitude of the deviation increases with radius, reflecting the cumulative influence of the halo gravitational potential, although it vanishes at very large $r$.
In the ZAMO case, the splitting between the prograde and retrograde branches is reduced compared to the static case, indicating that part of the observed asymmetry originates from the local fluid rotation encoded in $(\omega(r)-\Omega)$. In contrast, in the $\Omega=0$ case the deviations are enhanced, highlighting the role of fluid rotational effects in shaping the orbital energetics.

The behavior of the angular momentum deviation, $\Delta L/L_{\rm Vac}\%$, shows a qualitatively different trend. Unlike the energy, $\Delta L/L_{\rm Vac}\%$ grows monotonically with radius and becomes positive in most configurations, implying that the environment increases the orbital angular momentum required to maintain circular motion. For ZAMO, the effect is slightly stronger for prograde orbits, where the deviations are mildly larger compared to the retrograde orbits. This difference possibly suggesting that frame dragging and halo effects may act in the same direction for prograde motion but partially counteract each other for retrograde motion. In the static case, however, differences are larger for retrograde orbits. 

These differences likely originate from the fluid motion relative to frame dragging. A moving point particle is influenced by both the spin of the BH and the gravitational field of the fluid. Depending on the fluid velocity, this leads to different modifications between prograde and retrograde shifts.
Overall, the figures demonstrate that the environment induces departures from vacuum dynamics, leading to a nontrivial modification of both orbital energy and angular momentum. The magnitude of these effects depends on the halo parameters and on the fluid rotation, providing a potential observational signature of halo-induced frame dragging in rotating spacetimes. These results establish how environmental matter and rotation jointly modify the characteristic orbital radii and constants of motion that govern strong-field dynamics near the black hole. The magnitude and radial dependence of the relative deviations depend explicitly on the fluid’s angular velocity prescription, indicating that environmental rotation leaves distinct imprints beyond those induced by the static matter distribution alone.

\section{Separability and epicyclic frequencies}\label{sec: results}

Having analyzed circular equatorial motion and its rotational corrections, we now turn to a more general investigation of geodesic dynamics in the slowly rotating background spacetime. In particular, we examine the separability of the geodesic
equations using the Hamilton-Jacobi formalism, before moving to the computation of the epicyclic frequencies. We assume that the Hamilton-Jacobi action $S$ for a test particle of unit rest mass admits the standard additive ansatz:
\begin{equation}
S = - E t + L \phi + S_r(r) + S_{\theta}(\theta),
\label{HJ_action}
\end{equation}
where $E$ and $L$ are constants of motion associated with the stationarity and axial symmetry of the spacetime, corresponding to the conserved energy and azimuthal angular momentum per unit mass of the moving particle, respectively. The functions $S_r(r)$ and $S_{\theta}(\theta)$ encode the radial and polar dynamics, while being independent of $t$ and $\phi$ due to the symmetry. Note that $S_r$ and $S_{\theta}$ have different functional dependence, which enforces the separability.

We want to find a solution of the action $S$ by imposing that it satisfies the relativistic Hamilton-Jacobi equation, 
\begin{equation}
g^{\mu\nu}\partial_{\mu} S \, \partial_{\nu} S = - 1 ,
\label{HJ_equation}
\end{equation}
where $g^{\mu\nu}$ denotes the inverse metric tensor. Such constraints help connect the momentum with the derivatives of the action. Substituting the ansatz \eqref{HJ_action} into Eq.~\eqref{HJ_equation}, we find that, to linear order in the rotation parameter, the Hamilton-Jacobi equation separates completely into radial and angular parts, as it does also for the non-spinning configuration.

Explicitly, one obtains,
\begin{align}
p_r^2 \equiv\left(\frac{dS_r}{dr}\right)^2 &=
\frac{e^{2\lambda_0}}{r^2}
\bigg[
- K
+ r^2 E^2 e^{-2\nu_0}
\nonumber\\&- 2 r^2 e^{-2\nu_0}\, E L \omega
- r^2
\bigg],
\label{Sr_eq}
\\
p_{\theta}^2\equiv\left(\frac{dS_{\theta}}{d\theta}\right)^2 &=
K - \frac{L^2}{\sin^2\theta},
\label{Stheta_eq}
\end{align}
where $K$ is a constant of separation. Using Eq.~\eqref{Stheta_eq}, the separation constant $K$ can be written in the manifestly positive-definite form,
\begin{equation}
K = p_{\theta}^2 + \frac{L^2}{\sin^2\theta},
\label{Carter_constant}
\end{equation}
which plays the role of a Carter-like constant for the present spacetime. In particular, for equatorial motion ($\theta = \pi/2$), this reduces to $K = L^2$. Remarkably, separability of the Hamilton–Jacobi equation is preserved at linear order in spin. Higher-order rotational corrections are likely to break this property in general.

In stationary and axisymmetric spacetimes, the motion of test particles on nearly circular equatorial orbits is characterized by three fundamental frequencies: the orbital (azimuthal) frequency $\nu_\phi$, the radial epicyclic frequency $\nu_r$, and the vertical (polar) epicyclic frequency $\nu_\theta$. In the Kerr spacetime, these frequencies are functions of the black hole mass and spin, with the radial epicyclic frequency vanishing at the ISCO.

Using the geodesic equations, we calculate the epicyclic frequencies governing small oscillations about circular equatorial orbits. Expanding the effective potential to second order in radial and vertical perturbations, we obtain the radial and vertical epicyclic frequencies. For equatorial circular motion, the radial $(\nu_r)$ and vertical $(\nu_{\theta})$ epicyclic frequencies are given by:
\begin{align}
e^{2\lambda_0}\,\nu_{r\pm}^2 &=
\frac{e^{2\nu_0}}{r}
\left(
r\nu_0'' - 2 r \nu_0'^2 + 3 \nu_0'
\right)\nonumber\\
&\quad
\pm
e^{\nu_0}\sqrt{r  \nu_0'}
\left[
\left(3 - 4 r \nu_0'\right)\omega'
+ r \omega''
\right],\\
\nu_{\theta\pm}^2 &=
 \frac{e^{2\nu_0} \nu_0'}{r}
\pm
e^{\nu_0}\sqrt{r  \nu_0'}\,\omega',
\label{eq: epicyclic frequencies}
\end{align}
where the upper (lower) sign corresponds to prograde (retrograde) motion. These expressions explicitly demonstrate how frame dragging modifies the frequencies through terms linear in the spin parameter, which enter with opposite signs for prograde and retrograde motion. In the vacuum and non-rotating limit $\omega \to 0$, they reduce smoothly to the Schwarzschild epicyclic frequencies, while for small but finite spin, they agree with the vacuum small spin results \cite{Abramowicz:2003rc,Urbancova:2019btk}, thereby providing a non-trivial consistency check of our analysis. 

\begin{figure*}
\centering

\subfloat[]{\includegraphics[width=0.46\textwidth]{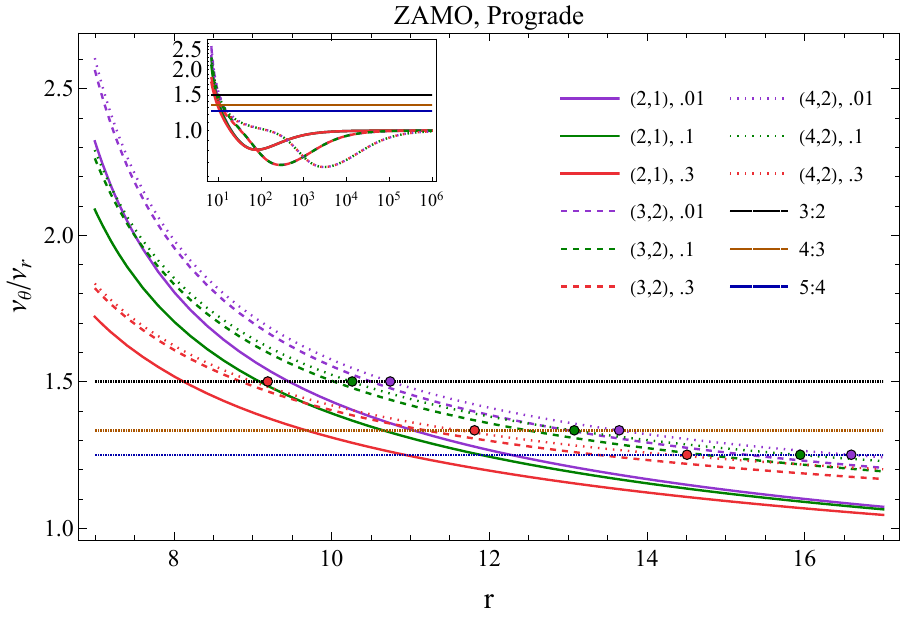}
\label{fig:ResCon_ZAMO_P}}
\hfill
\subfloat[]{\includegraphics[width=0.46\textwidth]{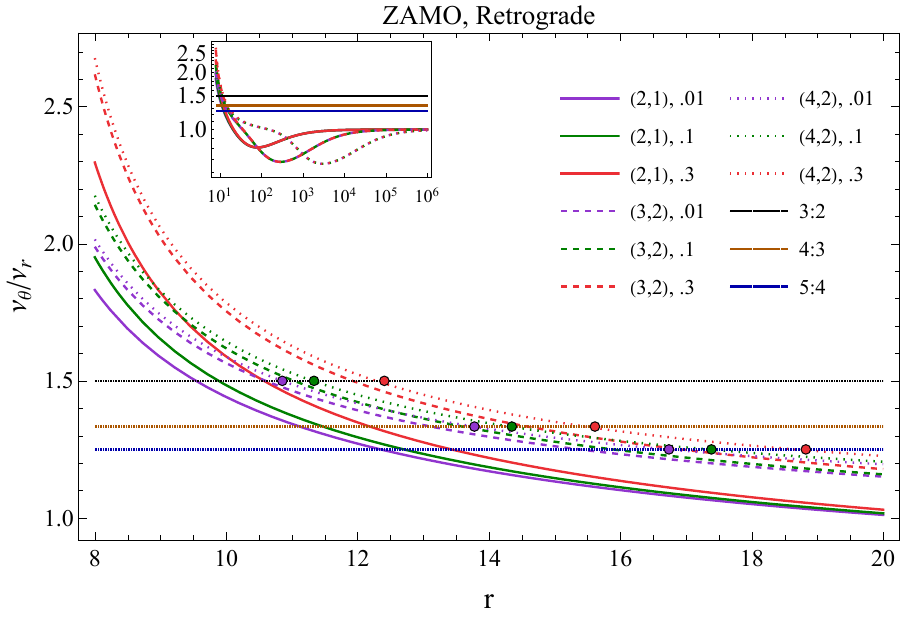}
\label{fig:Res_ConZAMO_R}}

\subfloat[]{\includegraphics[width=0.46\textwidth]{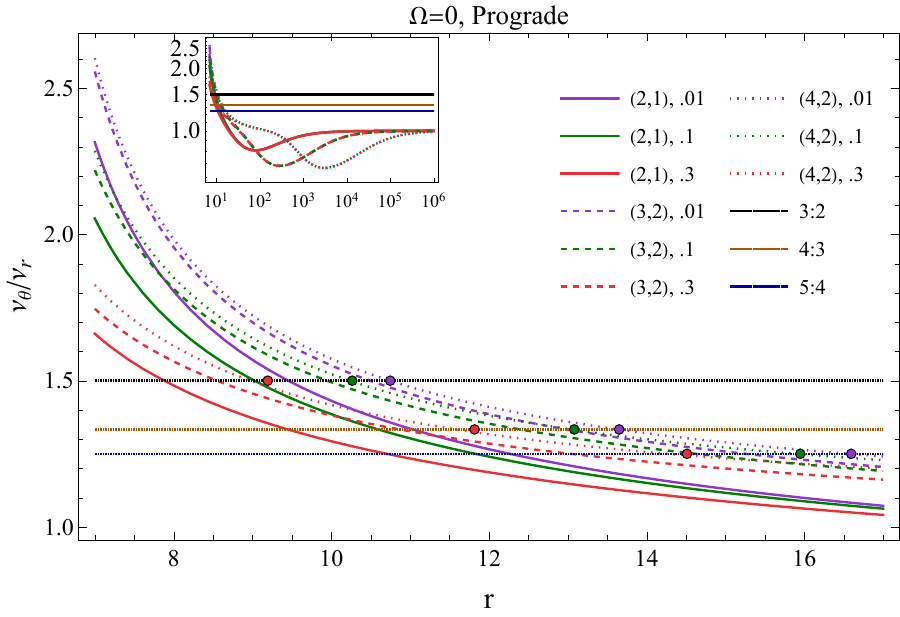}
\label{fig:ResCon_Ω0_P}}
\hfill
\subfloat[]{\includegraphics[width=0.46\textwidth]{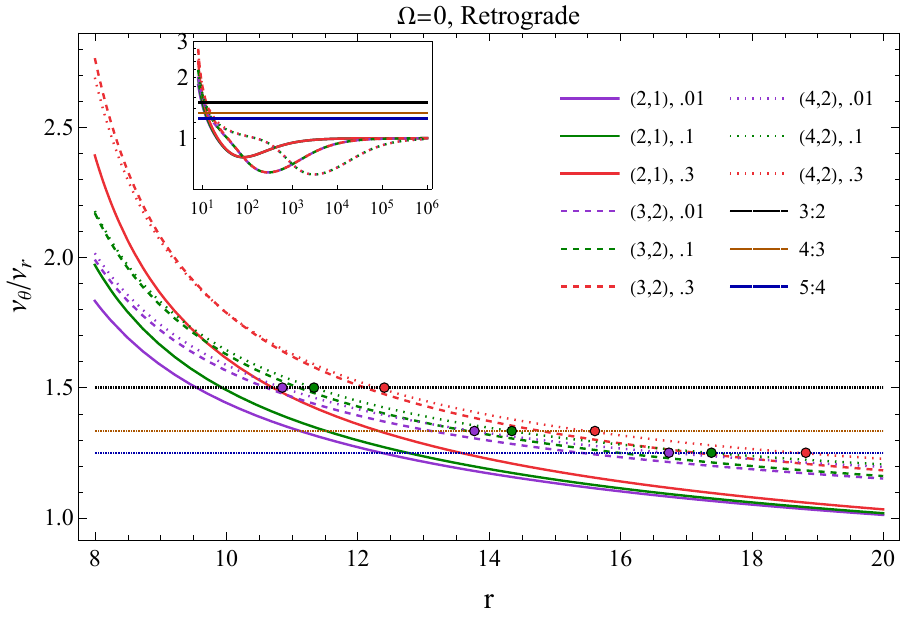}}
\label{fig:ResCon_Ω0_R}

\caption{Ratio of vertical to radial epicyclic frequencies, $\nu_\theta/\nu_r$, as a function of radius. Left (right) panels correspond to prograde (retrograde) motion. Upper panels show the ZAMO prescription, lower panels the static fluid ($\Omega=0$). Purple, sky blue, green, and red colors denote spin values $\chi=0.01,0.1,0.3$, and Solid, dashed, and dotted line styles indicate halo configurations $(2,1)$, $(3,2)$, and $(4,2)$. Horizontal lines mark the $3\!:\!2$, $4\!:\!3$, and $5\!:\!4$ resonances. Colored disks indicate the corresponding vacuum resonance positions. Environmental rotation induces systematic shifts and introduces a nonmonotonic structure in the frequency ratio. While different halo configurations produce nearly degenerate results at small spin, the separation between ZAMO and static fluids becomes increasingly pronounced at $\chi=0.3$, carrying physically distinct information about the fluid rotation. The explicit radial positions of the corresponding resonances are provided in Table~\ref{tab:Resonance_radii}.}
\label{fig:resonances}
\end{figure*}

Of particular interest are radii at which the epicyclic frequencies satisfy simple integer ratios,
\begin{equation}
\frac{\nu_\theta}{\nu_r} = \frac{p}{q}, \qquad p,q \in \mathbb{Z},
\end{equation}
which defines resonance conditions between vertical and radial oscillations. These resonances play a central role in the epicyclic resonance model for high-frequency quasi-periodic oscillations (HFQPOs) observed in accreting black hole and neutron star systems \cite{Abramowicz:2001bi,Torok2005}. 

In EMRIs around vacuum black holes, resonant crossings are known to induce transient non-adiabatic effects that can lead to cumulative waveform phase corrections \cite{Flanagan:2010cd, Ruangsri:2013hra, Berry:2016bit, Speri:2021psr, Gupta:2022fbe, Isoyama:2021jjd}. Over the large number of orbital cycles characterizing an EMRI, encounters with low-order resonances are expected to be generic \cite{Apostolatos:2009vu, Brink:2013nna, Brink:2015roa, Mukherjee:2019jhd}. Environmental perturbations further complicate this picture and may lead to additional structure, including resonant islands \cite{Destounis:2025tjn, Katagiri:2026gkz}. It is therefore necessary to quantify how environmental rotation shifts resonance radii relative to their vacuum values.

In vacuum BH spacetimes, the resonance radii are uniquely determined by the black hole parameters. However, due to the presence of the environment, the radial locations $r_{p:q}$ at which the resonance condition is satisfied are shifted relative to their vacuum values,
\begin{equation}
r_{p:q} = r_{p:q}^{\rm Vac} + \delta r_{p:q}^{\rm env},
\end{equation}
where $\delta r_{p:q}^{\rm env}$ encodes the influence of the surrounding matter distribution.

The study of these shifts is particularly important in the context of precision tests of strong gravity using EMRIs and HFQPOs. Since HFQPOs probe the innermost regions of accretion flows, environmental corrections to the epicyclic frequencies may lead to measurable deviations in the inferred resonance radii and, consequently, in estimates of BH mass and spin. Therefore, quantifying how environmental effects displace the $\nu_\theta$-$\nu_r$ resonances provides a powerful tool to disentangle genuine strong-field signatures from astrophysical systematics and to assess the robustness of black hole parameter measurements in non-vacuum spacetimes.


On the other hand, resonances in EMRIs break the adiabatic approximation commonly employed in EMRI waveform modeling. When the system passes through a resonance, previously averaged self-force effects can coherently accumulate, producing a sudden ``kick'' in the orbital constants of motion. This results in a dephasing of the gravitational waveform relative to adiabatic predictions, with potentially observable consequences for long-duration signals \cite{Flanagan:2010cd, Warburton2012}. In realistic astrophysical settings, environments perturb the background spacetime and modify the epicyclic frequencies. As a result, the inspiral encounters the resonance at a different orbital radius or time than predicted by vacuum models.

\begin{table*}
\centering
\setlength{\tabcolsep}{3pt}
\renewcommand{\arraystretch}{0.9}


\begin{minipage}{0.49\textwidth}
\centering
\text{Prograde (ZAMO)}\\[3pt]
\resizebox{.6\textwidth}{!}{
\begin{tabular}{ccccc}
\hline
Profile & $\chi$ & $r_{3:2}$ & $r_{4:3}$ & $r_{5:4}$ \\
\hline
(2,1) & 0.01 & 9.46 & 11.04 & 12.27 \\
(3,2) & 0.01 & 10.50 & 13.00 & 15.26 \\
(4,2) & 0.01 & 10.74 & 13.64 & 16.58 \\
Vac   & 0.01 & 10.74 & 13.64 & 16.58 \\[3pt]
(2,1) & 0.1  & 9.08 & 10.65 & 11.89\\
(3,2) & 0.1  & 10.03 & 12.48 & 14.72 \\
(4,2) & 0.1  & 10.24 & 13.06 & 15.91 \\
Vac   & 0.1  & 10.24 & 13.06 & 15.91 \\[3pt]
(2,1) & 0.3  & 8.12 & 9.68 & 10.92 \\
(3,2) & 0.3  & 8.89 & 11.19 & 13.37 \\
(4,2) & 0.3  & 9.00 & 11.62 & 14.30 \\
Vac   & 0.3  & 9.17 & 11.75 & 14.51 \\
\hline
\end{tabular}
}
\end{minipage}
\hfill
\begin{minipage}{0.49\textwidth}
\centering
\text{Retrograde (ZAMO)}\\[3pt]
\resizebox{.6\textwidth}{!}{
\begin{tabular}{ccccc}
\hline
Profile & $\chi$ & $r_{3:2}$ & $r_{4:3}$ & $r_{5:4}$ \\
\hline
(2,1) & 0.01 & 9.54 & 11.12 & 12.35 \\
(3,2) & 0.01 & 10.60 & 13.11 & 15.38\\
(4,2) & 0.01 & 10.85 & 13.77 & 16.72 \\
Vac   & 0.01 & 10.84 & 13.76 & 16.71 \\[3pt]
(2,1) & 0.1  & 9.88 & 11.47 & 12.70 \\
(3,2) & 0.1  & 11.03 & 13.59 & 15.88 \\
(4,2) & 0.1  & 11.32 & 14.32 & 17.35 \\
Vac   & 0.1  & 11.32 & 14.33 & 17.36 \\[3pt]
(2,1) & 0.3  & 10.57 & 12.19 & 13.43 \\
(3,2) & 0.3  & 11.92 & 14.58 & 16.92 \\
(4,2) & 0.3  & 12.28 & 15.47 & 18.66 \\
Vac   & 0.3  & 12.41 & 15.55 & 18.81 \\
\hline
\end{tabular}
}
\end{minipage}

\vspace{1em}


\begin{minipage}{0.49\textwidth}
\centering
\text{Prograde (Static)}\\[3pt]
\resizebox{.6\textwidth}{!}{
\begin{tabular}{ccccc}
\hline
Profile & $\chi$ & $r_{3:2}$ & $r_{4:3}$ & $r_{5:4}$ \\
\hline
(2,1) & 0.01 & 9.45 & 11.03 & 12.26 \\
(3,2) & 0.01 & 10.49 & 12.99 & 15.25 \\
(4,2) & 0.01 & 10.74 & 13.64 & 16.58 \\
Vac   & 0.01 & 10.74 & 13.64 & 16.58 \\[3pt]
(2,1) & 0.1  & 9.02 & 10.59 & 11.83 \\
(3,2) & 0.1  & 9.95 & 12.39 & 14.63 \\
(4,2) & 0.1  & 10.24 & 13.05 & 15.90 \\
Vac   & 0.1  & 10.24 & 13.06 & 15.91 \\[3pt]
(2,1) & 0.3  & 7.88 & 9.45 & 10.70 \\
(3,2) & 0.3  & 8.55 & 10.87 & 13.03 \\
(4,2) & 0.3  & 8.97 & 11.59 & 14.27 \\
Vac   & 0.3  & 9.17 & 11.75 & 14.51 \\
\hline
\end{tabular}
}
\end{minipage}
\hfill
\begin{minipage}{0.49\textwidth}
\centering
\text{Retrograde (Static)}\\[3pt]
\resizebox{.6\textwidth}{!}{
\begin{tabular}{ccccc}
\hline
Profile & $\chi$ & $r_{3:2}$ & $r_{4:3}$ & $r_{5:4}$ \\
\hline
(2,1) & 0.01 & 9.54 & 11.12 & 12.36 \\
(3,2) & 0.01 & 10.60 & 13.12 & 15.39 \\
(4,2) & 0.01 & 10.85 & 13.77 & 16.72 \\
Vac   & 0.01 & 10.84 & 13.76 & 16.71 \\[3pt]
(2,1) & 0.1  & 9.93 & 11.52 & 12.75 \\
(3,2) & 0.1  & 11.10 & 13.67 & 15.96 \\
(4,2) & 0.1  & 11.32 & 14.33 & 17.36 \\
Vac   & 0.1  & 11.32 & 14.33 & 17.36 \\[3pt]
(2,1) & 0.3  & 10.71 & 12.32 & 13.56 \\
(3,2) & 0.3  & 12.11 & 14.79 & 17.14 \\
(4,2) & 0.3  & 12.30 & 15.49 & 18.69 \\
Vac   & 0.3  & 12.41 & 15.55 & 18.81 \\
\hline
\end{tabular}
}
\end{minipage}

\caption{Resonance radii position is exhibited in the tables. $r_{p:q}$ represents the radial position of the $p:q$ resonances. We show ZAMO (static) fluid in the top (bottom) panel. The prograde (retrograde) orbits are shown in the left (right) panels. We also report the vaccum result (Vac) for comparison.}
\label{tab:Resonance_radii}
\end{table*}

Shifts in the resonance radius can lead to phase differences in EMRI waveforms. Consequently, environmental effects on epicyclic resonances represent a potential source of systematic error in EMRI parameter estimation if not properly accounted for. Conversely, precise measurements of resonance-induced waveform features may provide a novel probe of the near-BH environment of massive BHs, complementing other strong-field tests of gravity. Therefore, understanding how environmental perturbations alter radial-vertical epicyclic resonances is essential for robust EMRI waveform modeling and for fully exploiting the scientific potential of future space-based gravitational wave observations.


In Fig. \ref{fig:resonances} we show the ratio of the frequencies with respect to the radius along with the resonance position for $3:2$, $4:3$, and $5:4$. In the left column, we show the results for prograde orbits, and in the right column, we show the results for the retrograde orbits. The ZAMO and the static configurations are shown respectively in the upper and lower panels. Purple,  green, and red represent $\chi =$ 0.01,  0.1, 0.3, respectively. Solid, dashed, and dotted line styles represent the profile configurations for $(2,1)$, $(3,2)$, and $(4,2)$. The horizontal black, brown, and blue lines represent  $3:2$, $4:3$, and $5:4$ resonance, respectively. The coloured disks on the resonance lines show the resonance position for the corresponding spin in the absence of an environment. The corresponding numbers are provided in the Table \ref{tab:Resonance_radii}. In the inset of Fig. \ref{fig:resonances} we show the frequency ratios for the entire range of radius.

In vacuum Schwarzschild spacetime, the ratio of vertical to radial epicyclic frequencies, \(\nu_\theta/\nu_r\), is a monotonic function of radius, and each rational ratio \(p\!:\!q\) corresponds to a unique resonance radius \(r_{p:q}^{\rm Vac}\). This structure provides the natural baseline for assessing environmental corrections. When the surrounding fluid is included, the epicyclic frequencies are modified leading to systematic shifts of the resonance radii. In all cases considered, the ordering and existence of epicyclic resonances are preserved, and no qualitative deformation of the resonance hierarchy is observed. 
However, a local minimum in the epicyclic frequency ratio $\nu_{\theta}/\nu_r$, absent in vacuum spacetimes \cite{Torok2005}, emerges in the presence of an environment. This represents a qualitatively new dynamical feature induced purely by environmental effects and cannot be reproduced by vacuum. 

Crucially, this feature persists even in the limit of vanishing spin, as can be seen in the inset that all the curves of different spins for a given profile $(i,j)$ nearly overlap. This indicates that it is primarily driven by the matter distribution. The extended matter distribution creates a gravitational potential whose influence on the orbital dynamics is encoded in the background metric functions $\nu_0(r)$ and $\lambda_0(r)$ through the mass function $m(r)$. This modifies the radial structure of both $\nu_r$ and $\nu_\theta$, producing the local minimum.
Notably, the location of the local minimum tracks the characteristic scale radius $a_0$ of the halo profile, specifically, it occurs at a radius $r\sim \mathcal{O}( a_0/10)$.
In all the configurations considered, the frequency ratio $\nu_\theta/\nu_r$ asymptotically approaches the corresponding vacuum result at large radii, as expected from the fact that the halo density profile~\eqref{eq: density-Hern} falls off sufficiently rapidly at large $r$ so that environmental corrections to the metric become negligible. This provides an important consistency check of the framework and confirms that the non-monotonic structure, including the local minimum, is a genuinely strong/intermediate-field feature of the environment.


In Table~\ref{tab:Resonance_radii} we list the radial locations of resonances and their dependence on the density profile, fluid velocity, and spin $\chi$. Vacuum results are included for each spin to enable direct comparison with the fluid case.
Deviations from the vacuum case are negligible at low spin ($\chi=0.01$), but become noticeable as $\chi$ increases. Among the profiles, $(2,1)$ shows the largest deviations, while $(4,2)$ remains closest to vacuum values, indicating that configurations with matter located closer to the BH and larger $\mathcal{C}$ produce stronger modifications than those with more extended distributions.
The ordering of resonance positions is preserved in all cases.

The differences between static and ZAMO fluid configurations are negligible at low spin ($\chi = 0.01$), where both yield nearly identical resonance radii and closely match the vacuum case. Thus the choice of fluid velocity has little impact for slowly spinning systems.
In prograde configurations, ZAMO radii are generally slightly larger than the corresponding static ones, while in the retrograde case the same trend is present, i.e., static radii are larger than the ZAMO ones, though slightly less pronounced.
At higher spin ($\chi = 0.3$), these differences become more evident across all resonance ratios. This behavior is consistent with the increasing importance of frame dragging: the ZAMO fluid, which locally corotates with the spacetime, departs more significantly from the static case as rotation strengthens. Consequently, the discrepancy between the two descriptions grows with $\chi$, leading to measurable shifts in resonance positions.

Comparing the position of a particular coloured disk in Fig. \ref{fig:resonances} as well as comparing from Table~\ref{tab:Resonance_radii} clearly exhibits that the environmental impact on the resonance position increases with the increasing spin, irrespective of the rotational model. A consistent splitting between prograde and retrograde resonance radii is observed. This difference originates from frame dragging, which enters the equations of motion with opposite signs depending on the orbital orientation. The surrounding environment modulates the magnitude of this prograde-retrograde separation but does not introduce qualitative differences between the two branches. 
Second, the shifts are strongest for $(2,1)$ cases (solid curves), across the spin values (different colors), due to the large compactness and halo being distributed closer to the BH.

These results imply that environmental shifts of resonance radii generically modify the timing of transient resonances in EMRIs, potentially contributing to cumulative waveform phase corrections.  If detectable, these effects may allow simultaneous constraints on black hole parameters and on the distribution and dynamics of the surrounding matter. Together, these features demonstrate that environmental effects can introduce qualitatively new structure in epicyclic dynamics while preserving the overall resonance hierarchy.

\section{Conclusion}\label{sec:conclusion}

In this work, we have developed a self-consistent description of a slowly rotating black hole embedded in an external matter distribution, motivated by the fact that astrophysical compact objects are inevitably influenced by their surrounding environments. Focusing on the presence of an environment, we constructed the corresponding spacetime geometry and systematically investigated how environmental effects modify the metric and its geometric properties. 
This work provides a controlled semi-analytic framework incorporating environmental angular velocity into a slow-rotation expansion complementary to the recent fully numerical treatment of Ref.~\cite{Fernandes:2025osu}. The two frameworks together provide complementary handles on the problem. The numerical approach captures the full nonlinear structure at arbitrary spin, while the present semi-analytic treatment offers physical transparency, explicit parameter dependence, and computational efficiency. A detailed quantitative comparison with fully numerical solutions would provide an important validation of the slow-rotation regime and will be explored in future work.

Using this modified geometry, we analyzed several key orbital and dynamical observables, including the light ring, the ISCO, and the radial and vertical epicyclic frequencies. Our results demonstrate that the presence of an environment induces nontrivial and characteristic corrections to these quantities, even in the slow-rotation regime. These corrections encode detailed information about the environmental distribution and lead to deviations from the standard vacuum black hole predictions, highlighting the importance of environmental effects in precision studies of strong gravity.

The implications are particularly relevant for EMRIs, where small conservative modifications can accumulate over $10^4\!-\!10^5$ orbital cycles into measurable waveform dephasing. The shifts in the ISCO and epicyclic frequencies identified here provide a concrete mechanism through which environmental rotation may introduce systematic corrections in precision waveform modeling.
The conservative frequency shifts identified here persist throughout the strong-field inspiral. From the deviations in the orbital frequencies shown in Secs. \ref{sec: orbits} and \ref{sec: results}, typical fractional corrections in the strong-field region are at the level $\delta\Omega_o/\Omega_o \sim 10^{-4}–10^{-3}$ for representative halo parameters in the slow-rotation regime. For illustration, a fractional orbital frequency correction of order $\delta\Omega_o/\Omega_o \sim 10^{-4}–10^{-3}$ sustained over $N \sim 10^4–10^5$ cycles would accumulate into a total phase difference $\Delta\Phi \sim N\delta\Omega_o/\Omega_o \sim \mathcal{O}(1–10)$ radians. While a full waveform analysis is beyond the scope of the present work, this scaling indicates that environmental rotation can represent a potentially measurable systematic in precision EMRI modeling even in the slow-spinning case.

We also studied radial-vertical epicyclic resonances. In all cases, the resonance structure remains qualitatively identical to that of vacuum BH, with monotonic frequency ratios in the near region and with uniquely defined resonance radii. Environmental effects induce systematic shifts of the resonance locations. In particular, the emergence of nonmonotonic structure in the epicyclic frequency ratio away from BH highlights that environmental rotation can introduce qualitatively new strong-field features beyond simple parameter shifts. The magnitude of these shifts depends on the rotational properties of the fluid. Depending on the fluid rotation properties, the prograde or retrograde resonance position gets differently shifted. These results demonstrate that epicyclic resonances are robust against realistic environmental perturbations while remaining sensitive probes of small deviations from vacuum geometry, particularly relevant for precision EMRI studies.

In summary, our results demonstrate that (i) environmental matter modifies strong-field geometry such as ISCO, light ring, and resonance positions, (ii) environmental rotation introduces profile-dependent frame-dragging corrections, and (iii) EMRIs provide a uniquely sensitive probe of these effects due to cumulative phase accumulation over long inspirals.
More broadly, the framework introduced here offers a systematic, extensible approach to studying black holes embedded in realistic astrophysical environments. While the present work is restricted to slow rotation and a specific class of density distributions, the methodology can be generalized to higher-order rotational corrections, alternative matter profiles including accreting baryonic matter, or other forms of environmental media. Such extensions will be crucial for enhancing the fidelity of black hole spacetime models and for fully leveraging the potential of upcoming gravitational wave observations as probes of both strong field gravity and the nature of the surrounding environments.

\section*{Acknowledgments}
We thank Vitor Cardoso, Andrea Maselli and Sumanta Chakraborty for useful discussions.
S.D. acknowledges financial support from MUR, PNRR - Missione 4 - Componente 2 - Investimento 1.2 - finanziato dall'Unione europea - NextGenerationEU (cod. id.: SOE2024\_0000167, CUP:D13C25000660001).

\appendix

\section{Energy and angular momentum}
\label{app:Energy and angular momentum}

\begin{figure*}[ht]
\centering

{\includegraphics[width=0.46\textwidth]{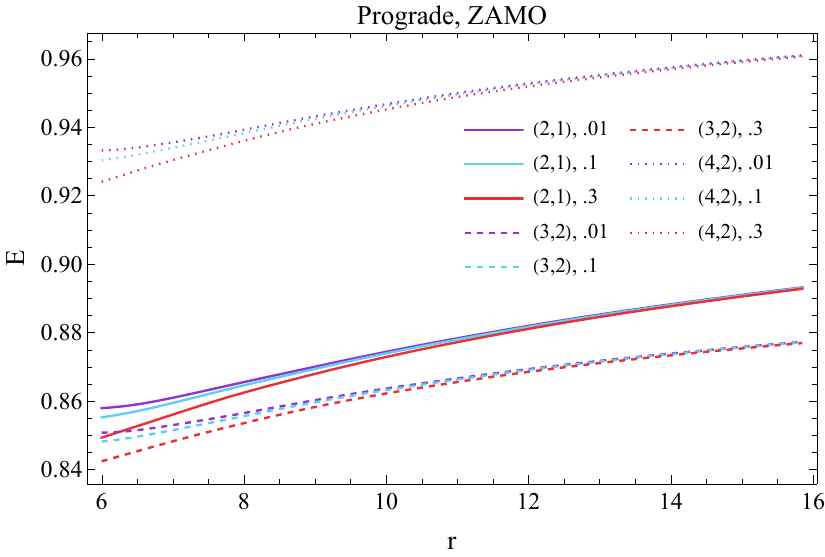}
\label{fig:}}
\hfill
{\includegraphics[width=0.46\textwidth]{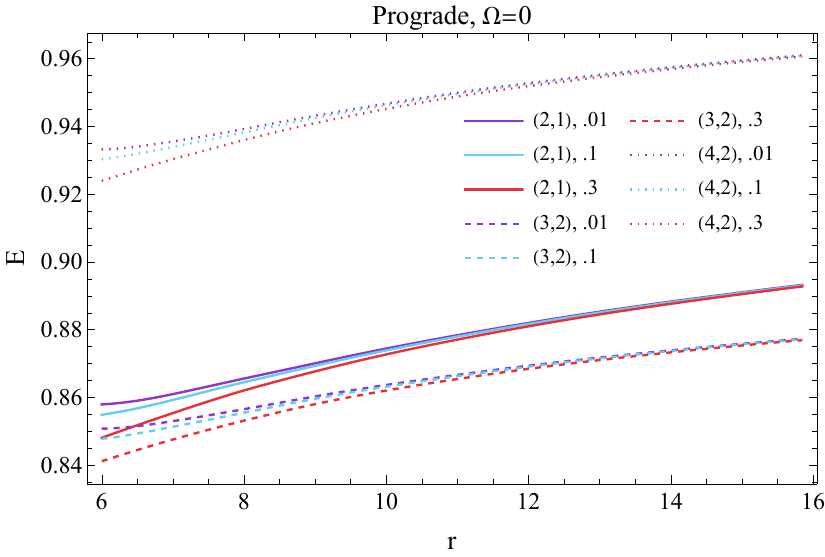}
\label{fig:}}

{\includegraphics[width=0.46\textwidth]{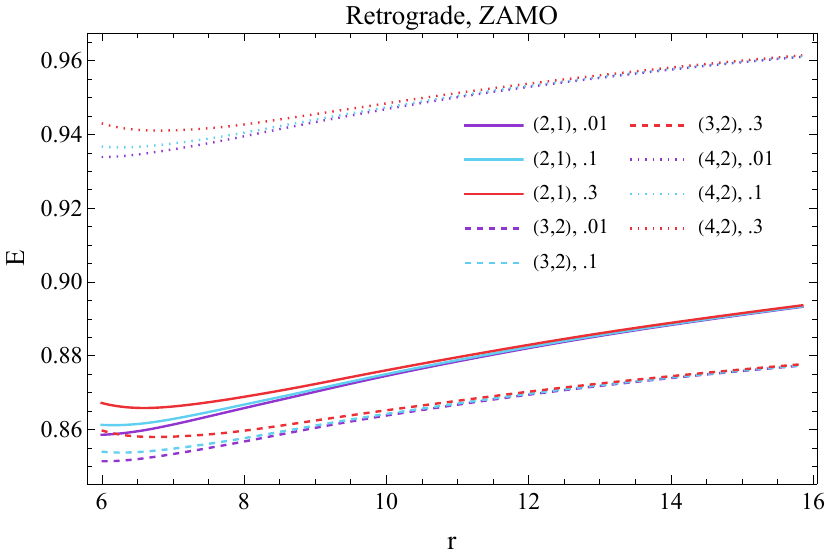}
\label{fig:}}
\hfill
{\includegraphics[width=0.46\textwidth]{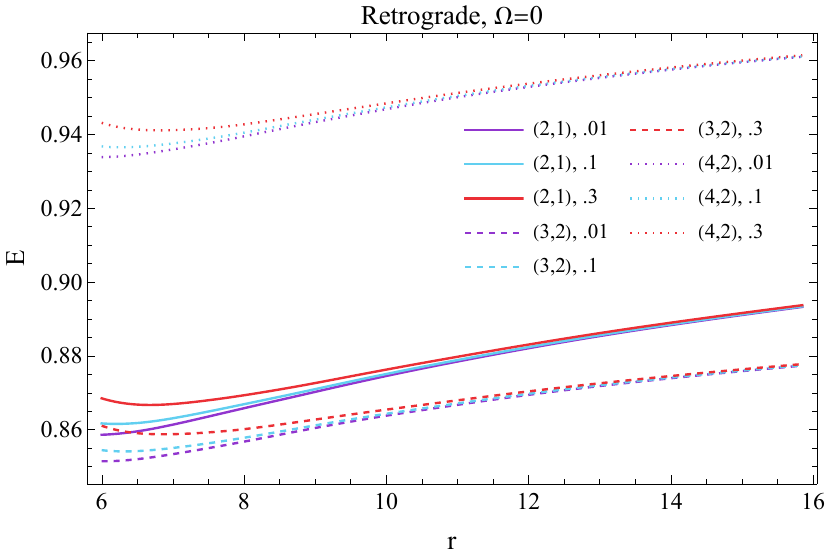}}
\label{fig:}

{\includegraphics[width=0.46\textwidth]{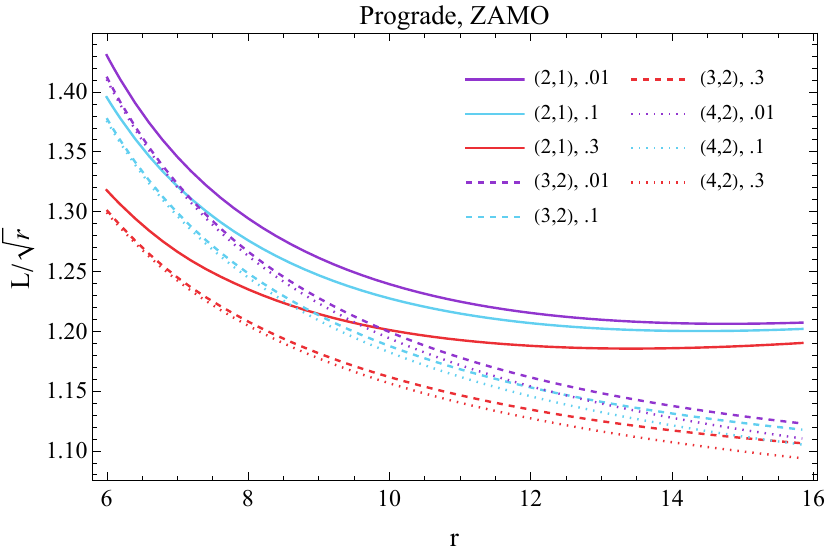}
\label{fig:}}
\hfill
{\includegraphics[width=0.46\textwidth]{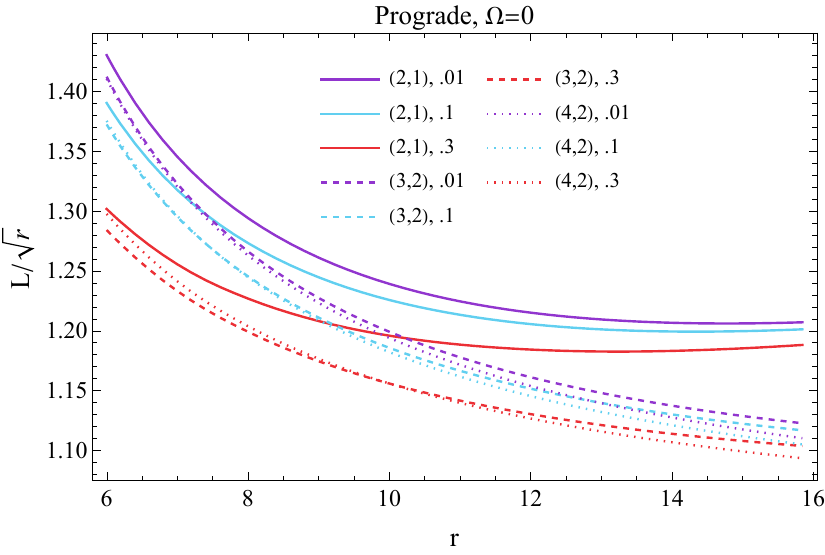}
\label{fig:}}

{\includegraphics[width=0.46\textwidth]{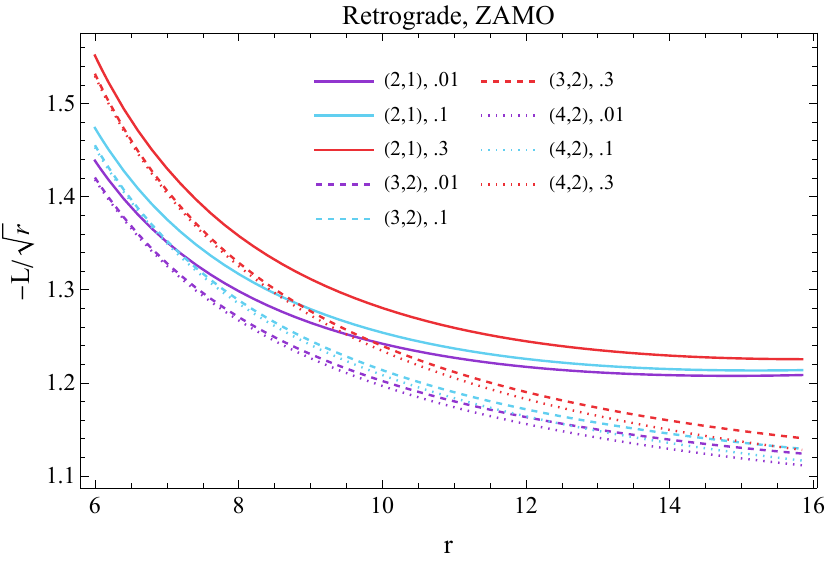}
\label{fig:}}
\hfill
{\includegraphics[width=0.46\textwidth]{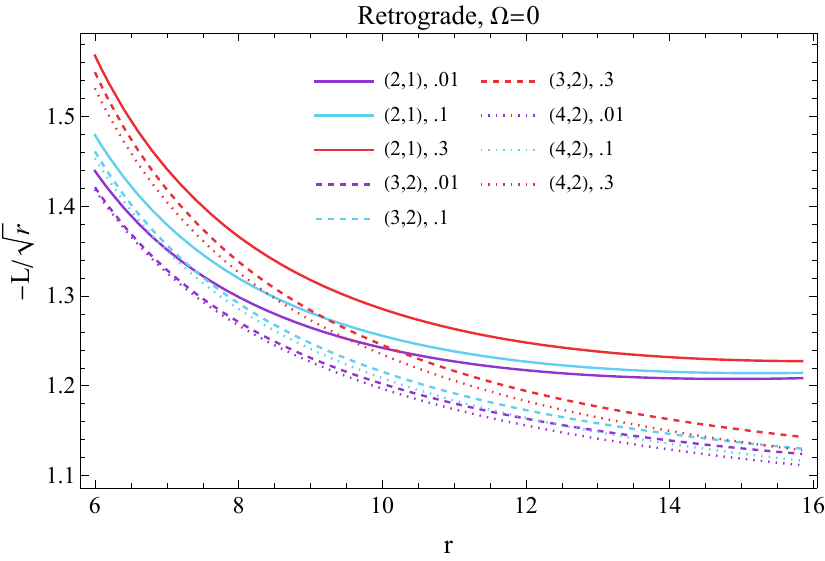}}
\label{fig:}

\caption{Conserved energy $E_\pm$ and angular momentum $L_\pm$ for equatorial circular timelike orbits. The first and third rows correspond to prograde motion and the second and fourth rows to retrograde motion. Left panels show the ZAMO prescription, right panels the static fluid. Deviations from the vacuum case arise from halo-induced modifications of the geometry and frame dragging.}
\label{fig:E-L all}
\end{figure*}

In this section we explicitly show the behavior of energy and angular momentum, that has been used to compute the fractional differences shown in Fig. \ref{fig:E change} and Fig. \ref{fig:L change}. In Fig. \ref{fig:E-L all} we show both the angular momentum and energy for all halo profiles and spins considered. The left column is for ZAMO and the right column is for static configuration. It shows the radial dependence of the specific energy \(E\) for equatorial circular orbits. In all cases, \(E\) increases monotonically with the orbital radius \(r\), reflecting the weakening of gravitational binding at larger distances, while approaching unity in the weak-field regime. The ordering of the curves is preserved throughout the radial range and is most
pronounced at small radii, where relativistic effects dominate.
As we are primarily interested in EMRI, we show energy and angular momentum for orbits that are in the near vicinity of the BH. Although $|L|$ decreases with increasing r in Fig \ref{fig:E-L all}, at $r\sim \mathcal{O}( a_0/10)$ it turns around and asymptotically at large $r$, $|L|/\sqrt{r} \sim \sqrt{(M_{\rm BH}+M_{\rm Halo})}$. Similarly asymptotically energy reaches $E\sim 1$.

Overall, slow rotation mainly introduces quantitative shifts between the different fluid motions and orbital orientations, whereas the density profile controls the depth of the effective potential and hence the binding energy of equatorial circular orbits.

\bibliography{references}

\bibliographystyle{apsrev4-1}

\end{document}